\documentclass[final]{article}
\usepackage{amsfonts, amsmath, amssymb, amsthm, mathtools}
\newtheorem*{thm}{Theorem}

\usepackage{fullpage}
\usepackage{authblk}

\usepackage{marvosym}
\usepackage{hyperref}

\usepackage[aboveskip=1pt,labelfont=bf,labelsep=period,justification=raggedright,singlelinecheck=off]{caption} 

\usepackage{tiaStyle}
\RequirePackage{listings}   

\definecolor{mygreen}{RGB}{28,172,0} % color values Red, Green, Blue
\DefineNamedColor{named}{myblue}{cmyk}{0.71,0.40,0,0.0}
\DefineNamedColor{named}{mygrey}{cmyk}{0.38,0.29,0.20,0.00}
\definecolor{mylilas}{RGB}{170,55,241}

\lstnewenvironment{cmatlab}
    {\lstset{language=Matlab,%
    breaklines=true,%
		basicstyle=\ttfamily\footnotesize\color{mygrey},
    morekeywords={matlab2tikz},
    keywordstyle=\color{myorange},%
    morekeywords=[2]{1}, keywordstyle=[2]{\color{black}},
    identifierstyle=\color{black},%
    stringstyle=\color{myblue},
    commentstyle=\color{mygrey},%
    showstringspaces=false,%without this there will be a symbol in the places where there is a space
    numbers=none,%
    numberstyle={\tiny \color{black}},% size of the numbers
    numbersep=9pt, % this defines how far the numbers are from the text
    emph=[1]{for,end,break},emphstyle=[1]\color{blue}, %some words to emphasise
		morekeywords={solve},
}}{}

\lstnewenvironment{matlab}
    {\lstset{language=Matlab,%
    breaklines=true,%
		basicstyle=\ttfamily\footnotesize\color{black},
    morekeywords={matlab2tikz},
    keywordstyle=\color{myorange},%
    morekeywords=[2]{1}, keywordstyle=[2]{\color{black}},
    identifierstyle=\color{black},%
    stringstyle=\color{myblue},
    commentstyle=\color{mygrey},%
    showstringspaces=false,%without this there will be a symbol in the places where there is a space
    numbers=left,%
    numberstyle={\tiny \color{black}},% size of the numbers
    numbersep=9pt, % this defines how far the numbers are from the text
    emph=[1]{for,end,break},emphstyle=[1]\color{blue}, %some words to emphasise
		morekeywords={solve},
}}{}

\title{Robust Parameter Estimation for Biological Systems: A Study on the Dynamics of Microbial Communities\footnote{This work was supported by the National Institute Of General Medical Sciences of the National Institutes of Health under Award Number R21GM107683.}}

\author[1]{Matthias Chung}
\author[2]{Justin Krueger}
\author[3]{Mihai Pop}

\affil[1]{Department of Mathematics, Virginia Tech, Blacksburg, VA, USA, \email{mcchung@vt.edu}}
\affil[2]{Department of Mathematics, Virginia Tech, Blacksburg, VA, USA, \email{kruegej2@vt.edu}}
\affil[3]{Center for Bioinformatics and Computational Biology, University of Maryland - College Park, College Park, Maryland, USA, \email{mpop@umiacs.umd.edu}}

\begin{document}
	
\maketitle

% \section*{abstract}
\begin{abstract}
	Interest in the study of in-host microbial communities has increased in recent years due to our improved understanding of the communities' significant role in host health. As a result, the ability to model these communities using differential equations, for example, and analyze the results has become increasingly relevant. The size of the models and limitations in data collection among many other considerations require that we develop new parameter estimation methods to address the challenges that arise when using traditional parameter estimation methods for models of these in-host microbial communities. In this work, we present the challenges that appear when applying traditional parameter estimation techniques to differential equation models of microbial communities, and we provide an original, alternative method to those techniques. We show the derivation of our method and how our method avoids the limitations of traditional techniques while including additional benefits. We also provide simulation studies to demonstrate our method's viability, the application of our method to a model of intestinal microbial communities to demonstrate the insights that can be gained from our method, and sample code to give readers the opportunity to apply our method to their own research.
\end{abstract}

\smallskip
\noindent \textbf{Keywords.} Parameter estimation, differential equations, Lotka-Volterra models, intestinal microbiota.

\section{Introduction}
The composition of in-host microbial communities (microbiota) plays a significant role in host health, and a better understanding of the microbiota's role in a host's transition from health to disease or vice versa could lead to novel medical treatments. One of the first steps toward this understanding is exploring the interaction dynamics of the microbes that compose the microbiota, which often are modeled using systems of differential equations. The size and complexity of microbiota dynamics, not to mention the difficulties involved in collecting sufficient data, makes this type of modeling exceedingly challenging. The inefficiencies and lack of robustness displayed by traditional parameter estimation techniques, such as single-shooting methods, only add to the challenge. Building on previously developed alternatives to traditional methods, we establish a novel parameter estimation method by approximating the model's state variables with spline functions and relaxing any known hard constraints on the model to achieve a new problem statement. This approach defines a ``nearby" parameter estimation problem that can be solved using robust numerical methods. We first verify our method on simulation studies using data generated by given generalized Lotka-Volterra equations. We then employ our method on data from an intestinal microbiota experiment, and we compare our results to a published parameterized model that uses the same data. In the simulation studies, we recover both the parameters and data, and in the comparison to the published intestinal microbiota model, our method exhibits superior data recovery. Our intestinal microbiota model also captures experimentally confirmed interactions. Based on these results, we conclude our robust method successfully parameterizes microbiota dynamics when modeled by a system of differential equations. This parameterization can lead to both qualitative and quantitative insights into the microbiota and direct future experiments to further improve the understanding of its dynamics.\\

This paper is structured as follows. In Sections~\ref{sub:biology} we provide biological background of the dynamics of microbial communities. A mathematical model of the interactions of microbial communities is introduced in Section~\ref{sub:lotkavolterra}. The standard setup for parameter estimation for ordinary differential equations is given in Section~\ref{sub:pe} followed by Section~\ref{sub:continuousshooting} on our novel parameter estimation method \emph{continuous shooting}. In Section~\ref{sec:results} we provide first a parameter estimation study for simulation data (Section~\ref{sub:simResults}) and secondly apply our continuous shooting method to a given data set on intestinal microbiota (Section~\ref{sub:microbiota}). Section~\ref{sec:dis} discusses our new methods and the results and implications from the microbiota data.

Note, to increase readability and quickly deliver our main points, we moved various details on our method, numerical derivations, and results to the appendix. Appendix~\ref{apx:splines} and~\ref{apx:splinesD} provide splines definitions and its required derivations. In Appendix~\ref{apx:comp} we give detailed information on the computations, simulation study, and the intestinal microbiota. A Matlab implementation of the main procedures is provided in Appendix~\ref{apx:matlab} including {\tt continuousShooting.m}, {\tt cubicSplines.m}, and {\tt lotkaVolterra.m}. Additionally a zip file is provided or can be downloaded at \href{http://www.math.vt.edu/people/mcchung/research.html}{\tt www.math.vt.edu/people/mcchung/} with all relevant Matlab codes to run an example problem (execute {\tt exampleScript.m}).

\subsection{Biology of Microbial Communities}\label{sub:biology}
Bacteria are ubiquitous in our world, and play a key role in maintaining the health of our environment as well as the health of virtually all living organisms. The human associated microbial communities (human microbiota) have been shown to be at least associated with, if not causative of, several human diseases such as periodontitis~\cite{Liu2012}, type 2 diabetes~\cite{Qin2012}, atopic dermatitis~\cite{Kong2012}, ulcerative colitis~\cite{Zella2010}, Crohn's disease~\cite{Macfarlane2009}, and vaginosis~\cite{Ravel2011}. Furthermore, time series data have shown that the host-associated microbiota undergo dynamic changes over time within a same individual, e.g., within the gut of a developing infant~\cite{Palmer2007,Koenig2011}, the gut, mouth, and skin of healthy adults~\cite{Caporaso2011}, and within the vagina of reproductive age women~\cite{Gajer2012}. The mechanisms that underlie these changes are currently not well understood, whether they represent fluctuations in the normal flora, or the transition from health to disease, and conversely from disease to health after treatment. 

Understanding the role human-associated microbial communities play in health and disease requires the elucidation of the complex networks of interactions between the microbes and between microbes and the host, a challenging task due to our inability to directly observe bacterial interactions. Instead, researchers have reconstructed microbial networks based on indirect approaches, such as knowledge about the metabolic functions encoded in the genomes of the interacting partners~\cite{Levy2013}, coexistence patterns across multiple samples~\cite{Chaffron2010}, covariance of abundance across samples~\cite{Friedman2012}, or changes in abundance across time~\cite{Ruan2006,Stein2013}. Multiple mathematical formalisms have been used to reason about the resulting networks with examples including metabolic modeling through flux balance analysis~\cite{Stolyar2007}, machine learning algorithms based on environmental parameters~\cite{Gilbert2011}, and differential equation based models of interactions~\cite{Stein2013}.  

Here we focus on the latter, a flexible formalism that can model complex interaction patterns, including abundance-dependent interaction parameters~\cite{Trosvik2008}. While such modeling approaches have been developed since the 1980s in the context of wastewater treatment systems~\cite{Henze1987}, their use in studying human-associated microbial communities has been limited, in no small part due to the specific characteristics of human microbiome data. First, the rate at which samples can be collected is severely limited by clinical and logistical factors, e.g., stool samples can be collected roughly on a daily basis, while subgingival plaque may only be feasibly collected at an interval of several months. Second, microbiome data are sparse, i.e., most organisms are undetected in most samples~\cite{Paulson2013} due to the detection limits of sequencing-based assays as well as the high variability of the microbiota across the human population.  Third, it is difficult if not impossible to directly measure environmental parameters, such as nutrient concentrations, that may impact the microbiota.

These features of the data derived from the human-associated microbiota lead to an ill-posed parameter estimation problem since multiple parameter sets may be consistent with the data. Numerical instabilities that result from specific parameter sets can also cause traditionally used estimation procedures to fail. Here we explore solutions to the parameter estimation problems in the context of the Lotka-Volterra formalism, which is described in more detail below.

\subsection{Mathematical Modeling of Microbial Communities}\label{sub:lotkavolterra}
We focus on a special type of differential equation based models of interactions, the Lotka-Volterra model, which is named after Alfred J. Lotka (1880-1949), an American mathematician, physical chemist, and statistician, and Vito Volterra (1860-1940), an Italian mathematician and physicist~\cite{Bacaer2011}.  This model was originally developed in the context of predator-prey interactions; however, it can be generalized to more complex interactions. Let $\bfy$ be the time dependent state variable for the dynamics of $n$ species with time variable $t$. Then the Lotka-Volterra system can be written as
\begin{equation} \label{eq:lv}
	\bfy' = \bff(\bfy) = \diag{\bfy} (\bfb + \bfA \bfy).
\end{equation}
Here, $\diag{\bfy}$ is the diagonal matrix with the state variables $\bfy = [y_1,\ldots,y_n]\t$ as its entries, and $\bfb = [b_1,\ldots,b_n]\t \in \bbR^n$ is the intrinsic growth rate, which incorporates the natural birth and death rate of each species in a given environment. Negative $b_i$ refers to a negative intrinsic growth rate and species $i$'s survival depends on the interaction with other species. The matrix $\bfA \in \bbR^{n\times n}$ represents the dynamics of the relationships between the species and is often referred to as the interaction matrix. An element $a_{ij} = [\bfA]_{i,j}$ of $\bfA$ describes the influence of species $j$ on the growth of species $i$. For $i \neq j$ and $a_{ij}<0$, we classically consider species $i$ to be a prey of predator $j$ and vice versa for $a_{ij}>0$. If for $i \neq j$ both $a_{ij}<0$ and $a_{ji}<0$, species $i$ and $j$ are competing for existence. On the other hand, if $i \neq j$ both $a_{ij}>0$ and $a_{ji}>0$, symbiotic behavior between species $i$ and $j$ is observed. If $a_{ij} = 0$, $i \neq j$, no direct interaction between species $i$ and $j$ exists. 

This formalism allows the simulation of ecological systems and the study of the long-term behavior of these systems. For example, the equilibrium solution $\bfy_\infty = \bfzero$ describes the extinction of all species. The equilibrium $\bfy_\infty = \bfzero$ is unstable if and only if at least one intrinsic growth rate $b_i$ is positive. All other biologically feasible, i.e., nonnegative, equilibrium solutions $\bfy_\infty$ of~\eqref{eq:lv} are solutions of the equation $\bf0 = \diag{\bfy}(\bfb + \bfA \bfy)$. The real parts of the eigenvalues of $\bff_\bfy(\bfy_\infty)$ determine the stability of the additional equilibria. Here, $\bff_\bfy(\bfy_\infty)$ is the Jacobian of $\bff$ with respect to $\bfy$ evaluated at $\bfy_\infty$ with $\bff_\bfy(\bfy) = \diag{\bfb}+\diag{\bfy}\bfA+\diag{\bfA\bfy}$. Detailed analysis on population dynamics, persistence, and stability can be found in~\cite{Sell2002}, and the specifics for the dynamics of Lotka-Volterra systems can be found in~\cite{Takeuchi1996}.

\subsection*{Assumptions and Prior Knowledge}
The Lotka-Volterra system makes simplifying assumptions about the underlying biological system, in particular it assumes that the interaction between microbes is constant in time and independent of the abundance of the interacting partners.  As a result, certain types of microbial interactions, e.g., quorum sensing, cannot be effectively modeled. For the sake of computational tractability, we restrict ourselves to this traditional definition of the Lotka-Volterra model, but extensions that allow more complex interaction modalities~\cite{Trosvik2008} can also be addressed by the computational parameter estimation framework described below.

The number of parameters of the Lotka-Volterra model is proportional to the square of the number of interacting partners, complicating the parameter estimation problem for complex datasets, especially when the number of samples is limited.  Prior knowledge about the system is often available and can be used to mitigate the complexity of parameter estimation. In the context of host-associated microbiota, this prior knowledge may include:
\begin{description}
	\item[Known Parameters.] Intrinsic growth rates or specific interactions might be known or partially known prior to parameter estimation. 
	\item[Grouping.] A reduction in the size of the system in~\eqref{eq:lv} can be achieved if species with similar behavior can be pooled together into a meta-species.  
	\item[Biomass.] Often a reasonable assumption is the total biomass in a dynamical system remains constant or is tightly regulated at all time.  
	\item[Symmetry.] Knowing the influence of species $j$ on species $i$ may simultaneously give information on both interaction parameters $a_{ij}$ and $a_{ji}$.
	\item[Finite Carrying Capacities.] It can be assumed that all species display logistic growth and have a finite carrying capacity in the absence of all other species.
	\item[Sparsity.] For some biological systems, an interaction between species or the lack thereof might be known. Even in cases when an exact sparsity pattern is not known, knowledge of interaction sparsity can be informative.
\end{description}

In this work we develop a novel parameter estimation method to recover intrinsic information of the dynamical system, i.e., the parameters, or predictors as they are called in statistics, given temporal density observations of the interacting species. In addition, this method is flexible enough to make use of any prior knowledge of the biological system such as the possible assumptions listed above. In particular, we propose the explicit incorporation of a sparsity assumption in the model.

\section{Methods}
\subsection{Problem Statement}\label{sub:pe}
In order to validate and impel model predictions, the mathematical model needs to be compared to experimentally observed data $\bfd \in \bbR^m$. The estimation of parameters for dynamical systems is a key step in the analysis of biological systems. The point estimates give quantitative information about the system, and in the case of a Lotka-Volterra system specifically, the intrinsic growth rates and interaction dynamics between species. Let us assume intrinsic growth rates $\bfb$ and interaction dynamics $\bfA$ are unknown and are collected in the parameter vector $\bfp = [\bfb; \vec{\bfA}]$, where $\vec{\bfA}$ is the concatenation of the columns of $\bfA$, i.e., $\vec{\bfA} = [a_{11},\ldots,a_{n1},a_{12},\ldots,a_{nn}]\t$. 

The general parameter estimation problem for any explicit first order ordinary differential equation (not just restricted to~\eqref{eq:lv}) is stated as the following constrained weighted least squares problem
\begin{equation}\label{eq:pe1}
\begin{gathered}
	 (\hat \bfp,\hat \bfy) = \argmin_{(\bfp,\bfy)} \norm[2]{\bfW(\bfm(\bfy(t;\bfp)) - \bfd)}^2 \\
	 \mbox{subject to} \quad \bfy' = \bff(t,\bfy,\bfp) \quad \mbox{and} \quad \bfc(\bfp,\bfy) = \bfzero, 
\end{gathered}	
\end{equation}
with $\bfy:[a,b]\times\bbR^n \to \bbR^n$. First, note that additional constraints such as prior knowledge discussed above are gathered by the general statement $\bfc(\bfp,\bfy) = \bfzero$. For every feasible parameter set $\bfp$ and initial state $\bfy_0 = \bfy(a)$ we assume that the conditions of the Theorem of Picard-Lindel\"{o}f are fulfilled. This means that a unique state variable $\bfy$ exists for any feasible choice $\bfp$ and $\bfy_0$. For our focus, the Lotka-Volterra system fulfills this condition for any finite $\bfp$ and $\bfy_0$. Further, $\bfm:\bbR^n \to \bbR^m$ is a projection from the state space onto the measurement space of given data $\bfd = [d_1,\ldots, d_m]\t \in \bbR^m$. For instance, observations $\bfd$ might not include all states at all time points, $\bfd$ might be in a frequency domain, or $\bfd$ may only be a combination of observed states. The precision matrix $\bfW\t\bfW$ is also referred to as the inverse covariance matrix, and the optimization problem~\eqref{eq:pe1} can be seen as a weighted least squares problem with weight matrix $\bfW$. Here, we assume that we have independent samples $\bfd$ and $\bfW = \diag{\bfw} = \diag{[w_1,\ldots,w_m]\t}$, with $w_j>0$ for $j = 1,\ldots,m$. If $w_j$ is large, observation $d_j$ plays an important role for the parameter estimation procedure. Otherwise, small $w_j$ indicate a lesser role for observation $d_j$. The underlying statistical assumption for this parameter estimation problem is that the residuals are normal distributed and in case of a diagonal $\bfW$ are uncorrelated~\cite{Calvetti2007}.

Computationally, solving~\eqref{eq:pe1} may be challenging. A limited number of observations, high levels of noise in the data,  large dynamical systems, non-linearity of the system, and a large number of unknown parameters are all such challenges. These challenges appear for Lotka-Volterra models of biological systems and make parameter estimation extremely difficult ~\cite{Aster2012}. Next, we note the methods traditionally used to solve the parameter estimation problem, discuss the limitations of these methods, and present a novel alternative. 

\subsection*{Traditional Methods}
Typically, single shooting methods are utilized to solve~\eqref{eq:pe1} for biological systems~\cite{Stoer2002,Bandara2009}. For single shooting methods, first, initial guesses for $\bfp^0$ and $\bfy_0^0$ are used to numerically solve the initial value problem (forward problem) using single- or multi-step methods such as Runge-Kutta and Adams-Bashforth methods~\cite{Hairer1993,Hairer2010}. Next, the misfit between data and model is computed, and depending on the optimization strategy (e.g., gradient based strategies such as Gauss-Newton methods or direct search approaches such as the Nelder-Mead Method), a new set $(\bfp^1, \bfy_0^1)$ is chosen. This process continues until a $\hat \bfp$ and $\hat \bfy_0$ are found to fulfill pre-defined optimality criteria. Since most efficient optimization methods are typically local optimization methods, globalization is achieved, for example, by Monte Carlo sampling of the search space, i.e., repeated local optimization with random initial guesses~\cite{Conrad2009}. The global minimizer chosen from the set of local minimizers is the local minimizer $(\hat \bfp, \hat \bfy_0)$ with minimal function value. Various other strategies for global optimization can be applied, such as simulated annealing~\cite{Kirkpatrick1983},  evolutionary algorithms~\cite{Simon2013}, or particle swarm optimization methods~\cite{Kennedy1995}. 

The main drawback of these methods, though, is the numerical forward solver often fails to integrate with the required precision meaning the calculations of the forward solution and the misfit between the data and model are not possible. It has been established that single shooting methods are not robust to initial guesses $\bfp^0$ and $\bfy_0^0$, which in this case refers to the methods' ability to successfully find minimizers as defined in~\eqref{eq:pe1}. Various alternatives have been developed to compensate for the lack of robustness~\cite{Liu2014, Bock1984}. In particular, multiple shooting methods divide the time interval in several smaller subintervals, introducing initial conditions for each subinterval and solves the individual initial value problems on each subinterval. Constrained optimization methods will ensure continuity of the optimized state variable~\cite{Bock1984} when using multiple shooting methods, so these methods introduce more robustness when compared to single shooting methods by the reformulating the problem as a constrained optimization. 

\subsection{Continuous Shooting}\label{sub:continuousshooting}
In this manuscript we follow another approach similarly discussed in~\cite{Poyton2006,Chung2012,Ramsay1996} by using an interpretation of a collocation-type method~\cite{Ascher1998}. Instead of solving the original optimization problem~\eqref{eq:pe1}, we approximate the elements of the state variable for the system by a function $s \in \calS$, where $\calS$ is an any function space dense in $\calC^1([a,b])$, such as the cubic spline space we use here. Furthermore, we relax~\eqref{eq:pe1} and use the standard approach of discretize-then-optimize~\cite{Nocedal2006,Vogel2002}.

We first reformulate~\eqref{eq:pe1} as 
\begin{equation}\label{eq:pe2}
	\begin{gathered}
	 \min_{(\bfp,\bfy)} \norm[2]{\bfW(\bfm(t;\bfy) - \bfd)}^2 \\
	 \mbox{subject to} \quad \norm[\calL^p]{\bfy' - \bff(t,\bfy,\bfp)}^p = 0 \quad \mbox{and} \quad \bfc(\bfp,\bfy) = \bfzero,
\end{gathered}
\end{equation}
where $\norm[\calL^p]{\,\cdot\,}$ is any appropriate integral norm with $p=2$ for our later examples. The problems defined in~\eqref{eq:pe1} and~\eqref{eq:pe2} are equivalent since $\bfy$ is required to be continuous. Choosing a distance metric $\calD$ and relaxing the optimization problem leads to 
\begin{equation*}\label{eq:pe3}
	\begin{gathered}
	 \min_{(\bfp,\bfy)} \norm[2]{\bfW(\bfm(t;\bfy) - \bfd)}^2 + \lambda \norm[\calL^2]{\bfy' - \bff(t,\bfy,\bfp)}^2 + \alpha \calD(\bfc(\bfp,\bfy)).
\end{gathered}
\end{equation*}
Here, $\lambda>0$ and $\alpha>0$ can be seen as either Lagrange multipliers or regularization parameters~\cite{Nocedal2006,Vogel2002}. The parameter $\lambda$ has the effect that if $\lambda$ is small, the main contributor to the minimization process is the data misfit. On the other hand, if $\lambda$ is large, the weight of contribution shifts to the model equations. A similar interpretation holds for $\alpha$. Note, further, that the conversion of the constraints $\bfc(\bfp,\bfy) = \bfzero$ to ``soft'' constraints is also optional. The constraints can remain as hard constraints and interior point methods or augmented Lagrangian methods can be utilized~\cite{Nocedal2006}.

Let $\calS^3_\bftau([a,b])$ be the set of cubic splines with knots $a = \tau_0< \cdots < \tau_k = b$ with a chosen set of boundary conditions, e.g., not-a-knot conditions, then every $s \in \calS^3_\bftau([a,b])$ is uniquely determined by a set of parameters $\tilde \bfq = [\tilde q_{0},\ldots,\tilde q_{k}]\t$. A vectorized spline $\bfs$ defined by $\bfs = [s_1,\ldots, s_n]\t$ with $s_j \in \calS^3_\bftau([a,b])$ for $j = 1,\ldots,n$ is then uniquely determined by a parameter vector $\bfq = [\tilde \bfq_1,\ldots,\tilde \bfq_n]\t$. Approximating the state variable $\bfy$ by $\bfs$, discretizing the integral of the $\calL^2$-norm, and normalizing $\lambda$ finally leads to the optimization problem
\begin{multline}\label{eq:pe4}
 \min_{(\bfp,\bfq)} \ \calJ(\bfp,\bfq) = \norm[2]{\bfW(\bfm(t;\bfs(t;\bfq)) - \bfd)}^2 + \frac{\lambda}{n\ell} \norm[2]{\bfs'(\bfT;\bfq) - \bff(\bfT,\bfs(\bfT;\bfq),\bfp)}^2  \\ + \alpha \calD({\bfc(\bfp,\bfs(t;\bfq)}),
\end{multline}
where we define
\begin{equation*}
\bfs'(\bfT;\bfq) = \begin{bmatrix*}[c]\bfs'(T_1;\bfq) \\ \vdots \\ \bfs'(T_\ell;\bfq)\end{bmatrix*},\quad \bff(\bfT,\bfs(\bfT; \bfq),\bfp)=\begin{bmatrix*}[c]\bff(T_1,\bfs(T_1;\bfq),\bfp) \\ \vdots \\ \bff(T_\ell,\bfs(T_\ell;\bfq),\bfp)\end{bmatrix*},
\end{equation*}
and $a = T_1 < \cdots < T_\ell = b$ is a discretization of the interval $[a,b]$. We refer to this method as continuous shooting.

\subsection*{Numerics}
To numerically solve~\eqref{eq:pe4} with respect to $\bfp$ and $\bfq$, we can utilize Gauss-Newton type methods. Suppose $\calD$ is the two-norm, so a standard optimization algorithm can be written as follows. Let the residuals of~\eqref{eq:pe4} be defined by
\begin{equation*}\label{eq:res}
\bfr = \begin{bmatrix*}
	\bfr_1 \\ \bfr_2 \\ \bfr_3
\end{bmatrix*} 
= 
\begin{bmatrix*}
	\bfW(\bfm(t;\bfs(t;\bfq))-\bfd) \\ 
	\sqrt{\lambda} (\bfs'(\bfT;\bfq) - \bff(\bfT,\bfs(\bfT;\bfq),\bfp)) \\ \sqrt{\alpha} \bfc(\bfp, \bfs(t;\bfq)). 
\end{bmatrix*}
\end{equation*}
and the Jacobian of $\bfr$ be given by
\begin{equation*}\label{eq:jac}
\bfJ = 
\begin{bmatrix*}
	\frac{\partial \bfr_1}{\partial \bfp} & \frac{\partial \bfr_1}{\partial \bfq} \\[1ex]
	\frac{\partial \bfr_2}{\partial \bfp} & \frac{\partial \bfr_2}{\partial \bfq} \\[1ex]
	\frac{\partial \bfr_3}{\partial \bfp} & \frac{\partial \bfr_3}{\partial \bfq}
\end{bmatrix*}
=
\begin{bmatrix*}
	\bfzero &  \bfW \bfm_\bfs  \bfs_\bfq \\[1ex]
	-\sqrt{\lambda} \bff_\bfp(\bfT) & \sqrt{\lambda} (\bfs_\bfq'(\bfT) - \bff_\bfs(\bfT) \bfs_\bfq(\bfT)) \\[1ex]
	\sqrt{\alpha} \bfc_\bfp & \sqrt{\alpha} \bfc_\bfs \bfs_\bfq
\end{bmatrix*}
\end{equation*}
with the appropriate abbreviations
\begin{equation*}
	\begin{aligned}
	\bfm_\bfs &= \frac{\partial}{\partial \bfs}\bfm(t;\bfs),\\
	\bfs_\bfq &= \frac{\partial}{\partial \bfq}\bfs(t;\bfq), \\
	\bfs_\bfq(\bfT) &= \frac{\partial}{\partial \bfq}\bfs(\bfT;\bfq), \\
	\bfs_\bfq'(\bfT) &= \frac{\partial}{\partial \bfq}\bfs'(\bfT;\bfq),
	\end{aligned}
	\hspace{20ex}
	\begin{aligned}
	\bff_\bfp(\bfT) &= \frac{\partial}{\partial \bfp}\bff(\bfT,\bfs,\bfp),\\
	\bff_\bfs(\bfT) &= \frac{\partial}{\partial \bfs}\bff(\bfT,\bfs,\bfp),\\	
	\bfc_\bfp &= \frac{\partial}{\partial \bfp}\bfc(\bfp,\bfq),\\		
	\bfc_\bfs &= \frac{\partial}{\partial \bfs}\bfc(\bfp,\bfs).	
	\end{aligned}	
\end{equation*}
Then the gradient of the objective function $\calJ$ is given by $\bfg = \nabla \calJ(\bfp,\bfq) = 2\bfJ\t \bfr$ and the Gauss-Newton approximation on the Hessian is $\bfH = \nabla^2 \calJ(\bfp, \bfq) \approx 2\bfJ\t\bfJ$. Finding the structures for $\bfm$, $\bfs$, $\bff$, $\bfc$, and their corresponding partial derivatives can be done analytically. The details for the construction of the cubic splines $\bfs$ and their partial derivatives can be found in Appendix~\ref{apx:splines} and~\ref{apx:splinesD}, and the details for the construction of $\bff$ and its partial derivatives when using the Lotka-Volterra system as a model can be found in Appendix~\ref{apx:comp}.

The final remaining piece that needs to be acknowledged in the problem defined by~\eqref{eq:pe4} is the determination of the regularization parameter(s) $\lambda$ and, if present, $\alpha$. The goal in choosing the regularization parameters $\lambda$ and $\alpha$ is to choose values such that the model system and any additional constraints are sufficiently weighted to prevent data overfitting but not overweighted to cause data underfitting. To accomplish this, we use $k$-fold cross-validation to select the regularization parameters~\cite{Geisser1979}. The reason we use cross-validation here is the method allows us to use a subset of known data to train our model by solving~\eqref{eq:pe4} and the remaining known data to test the model's predictive abilities. This is particularly useful in biological applications such as the ones being discussed here because the available data are often limited and hard to collect. 
 
\begin{figure}[h]
\includegraphics[width=\textwidth]{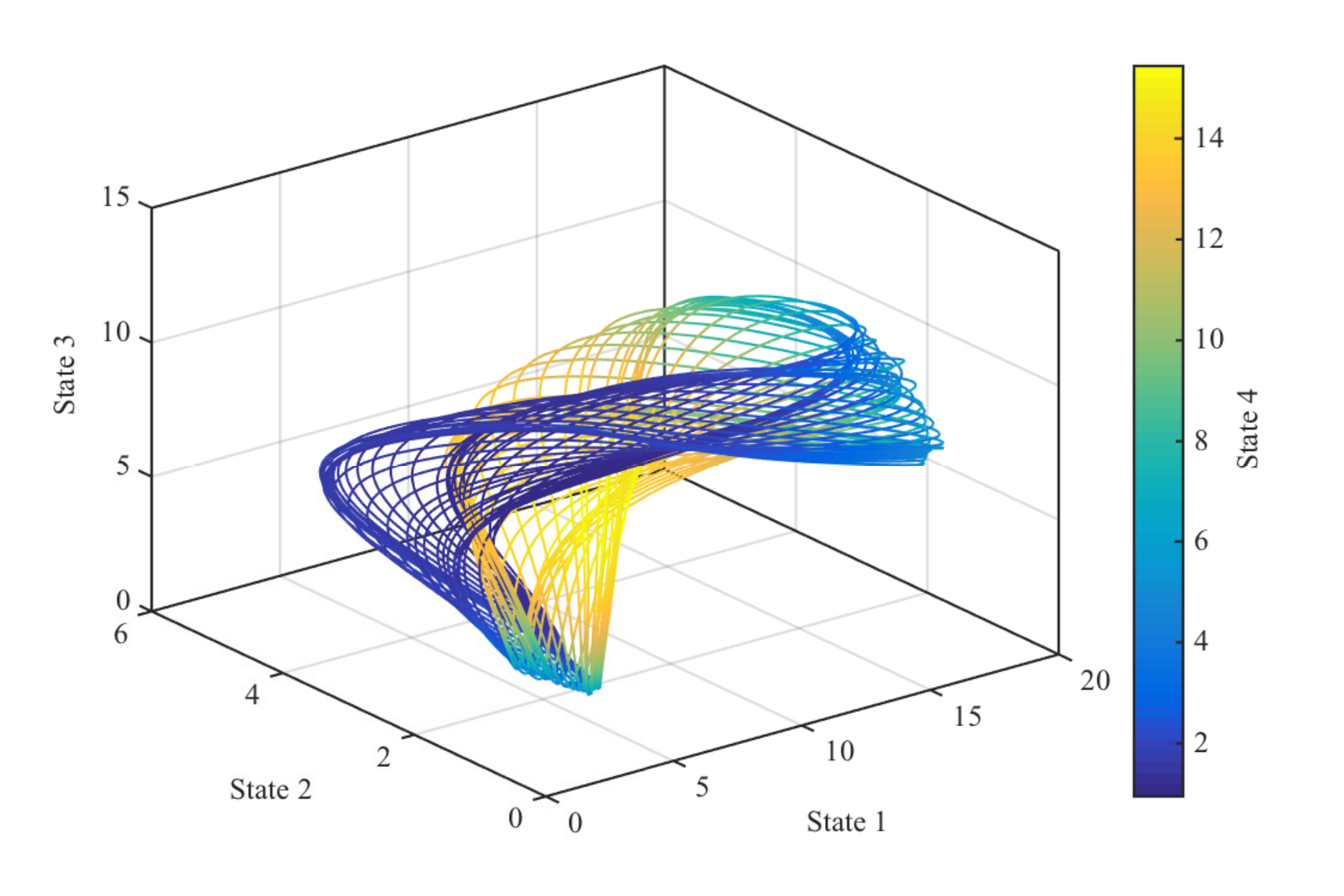}
\caption{{\bf Extended numerical solution of a Lotka-Volterra system.} \\ The plotted solution is from time $t = 0$ to $t = 200$ and demonstrates the chaotic behavior of the system.}
\label{fig:Fig1}
\end{figure}

\section{Results}\label{sec:results}

In this section we share our findings with regard to our method's viability and its applicability using simulation studies and previously collected and published data for intestinal microbiota, respectively. In both sets of findings, we focused on the Lotka-Volterra system of differential equations as a model and an unknown sparsity pattern in the interaction matrix as an assumption on the model.

\subsection{Simulation Studies}\label{sub:simResults}
On the interval $t \in [0,10]$, we defined a four dimensional Lotka-Volterra system using the parameters and initial conditions
\begin{equation*}
\bfb = \begin{bmatrix*}[r] 2 \\ 1\\ 0 \\ -3 \end{bmatrix*}, \quad \bfA = \begin{bmatrix*}[r] 0 & -0.6 & 0 & -0.2 \\ 0.6 & 0 & -0.6 & -0.2 \\ 0 & 0.6 & 0 & -0.2 \\ 0.2 & 0.2 & 0.2 & 0 \end{bmatrix*}, \quad \mbox{and} \quad \bfy_0 = \begin{bmatrix*}[r] 5 \\ 4 \\ 3 \\ 2 \end{bmatrix*}.
\end{equation*}
In this case, the interaction matrix displayed 37.5 percent sparsity. Using the parameters and initial conditions, we numerically solved the initial value problem with the differential equation given by~\eqref{eq:lv}. In Figure~\ref{fig:Fig1} a phase plot of the solutions for States 1 through 4 indicates the system displayed chaotic dynamics, and it is inherently difficult for parameter estimation methods to find parameters of chaotic systems.  

We then used the numerical solution to generate three sets of data with different levels of multiplicative noise (Study 1: 0 percent noise; Study 2: up to 10 percent noise; Study 3: up to 25 percent noise) and applied our parameter estimation method using each data set. The data for each study are shown in Figure~\ref{fig:Fig2}, and the remaining details for the problem setup are included in Appendix~\ref{apx:simu}. The code provided in Appendix~\ref{apx:matlab} also allows for setting up similar studies. 
\begin{figure}[h]
\includegraphics[width=\textwidth]{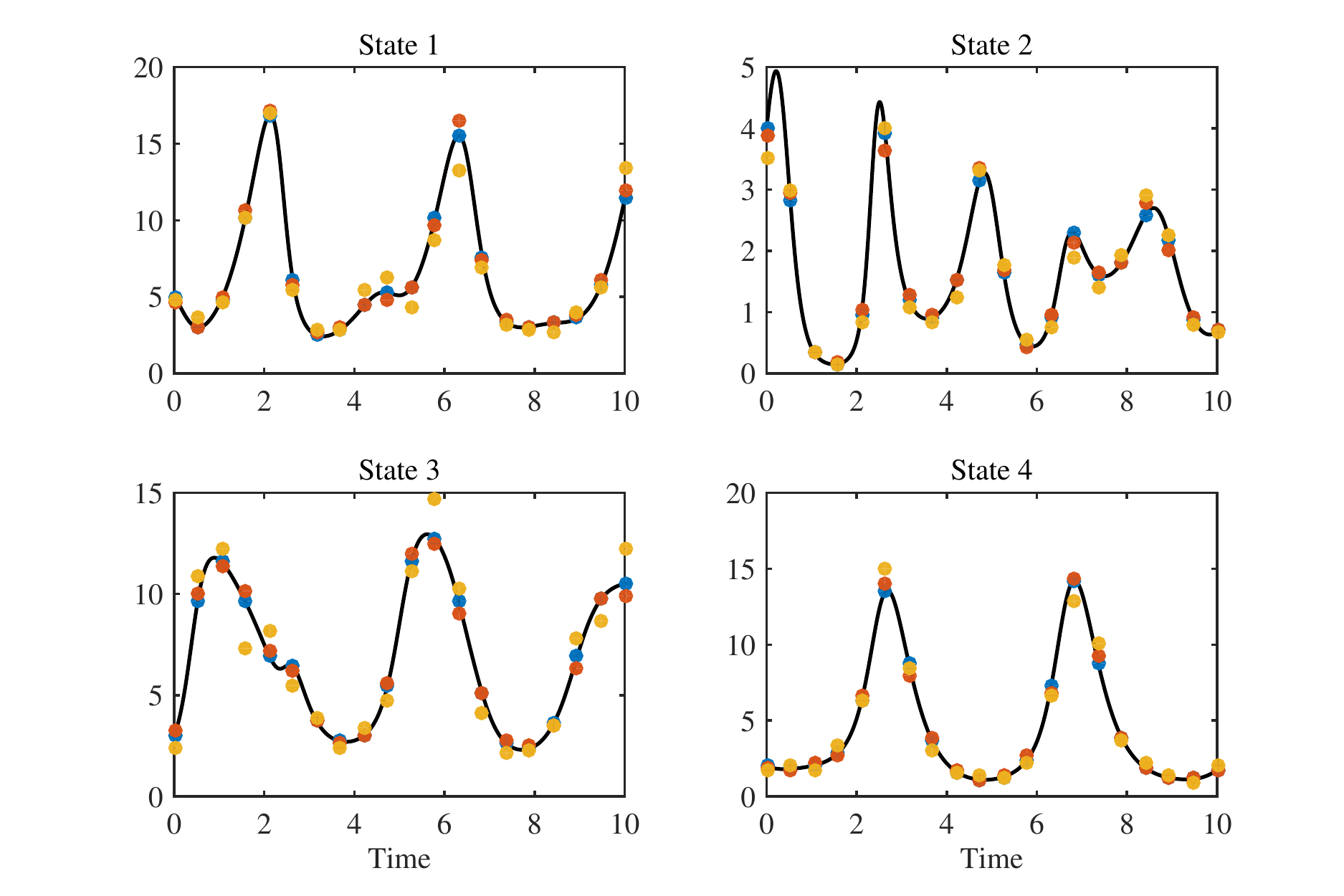}
\caption{{\bf Numerical solution of a Lotka-Volterra system and data points for simulation studies.} \\ The black curves indicate the numerical solutions used to generate the data. The blue dots are the data with no multiplicative noise. The red dots are the data with up to 10 percent multiplicative noise, and the yellow dots are the data with up to 25 percent multiplicative noise.}
\label{fig:Fig2}
\end{figure}

The state solutions of the system found using the optimal model parameters recovered the data very well in all three studies. The relative errors, defined as $e_{r} = \frac{1}{m}\norm[1]{\frac{\bfm(\bfy)-\bfd}{\bfd}}$ where $m$ is the number of elements in $\bfd$ and the division is element-wise, were approximately $e_{r} \approx 0.0331$, $e_{r} \approx 0.0926$, and $e_{r} \approx 0.1511$ for the three studies, respectively.

\begin{figure}[h]
\includegraphics[width=\textwidth]{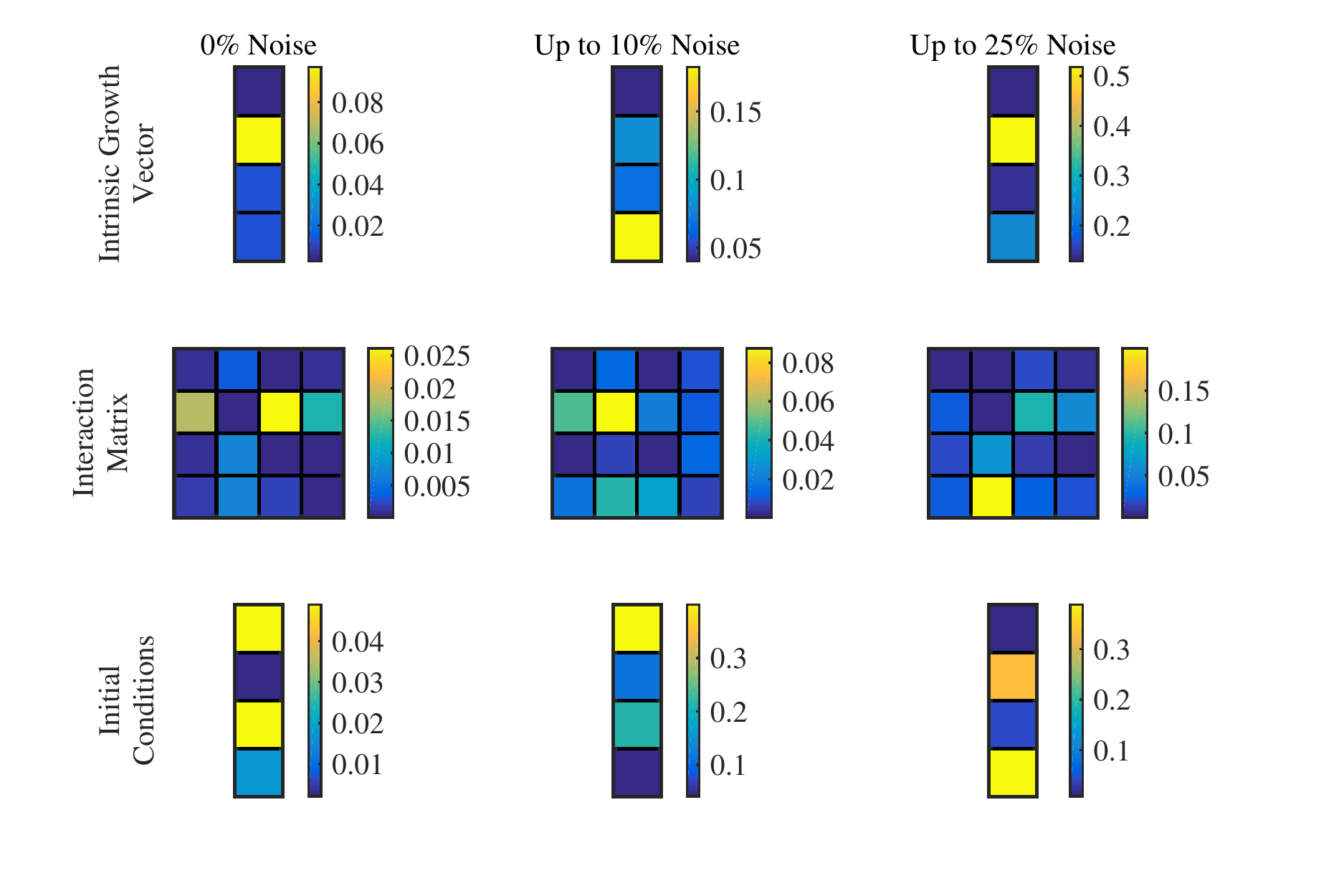}
\caption{{\bf Absolute errors of optimal model parameters and initial conditions.} \\ The images show the absolute errors in the recovery of the model parameters and initial conditions for all three simulation studies.}
\label{fig:Fig3}
\end{figure}

As Figure~\ref{fig:Fig3} demonstrates, we also compared our optimal model parameters to their true values, and the absolute errors in the optimal model parameters suggested the true model parameters and the system dynamics were recovered very well in all three studies.

\begin{figure}[h]
\includegraphics[width=\textwidth]{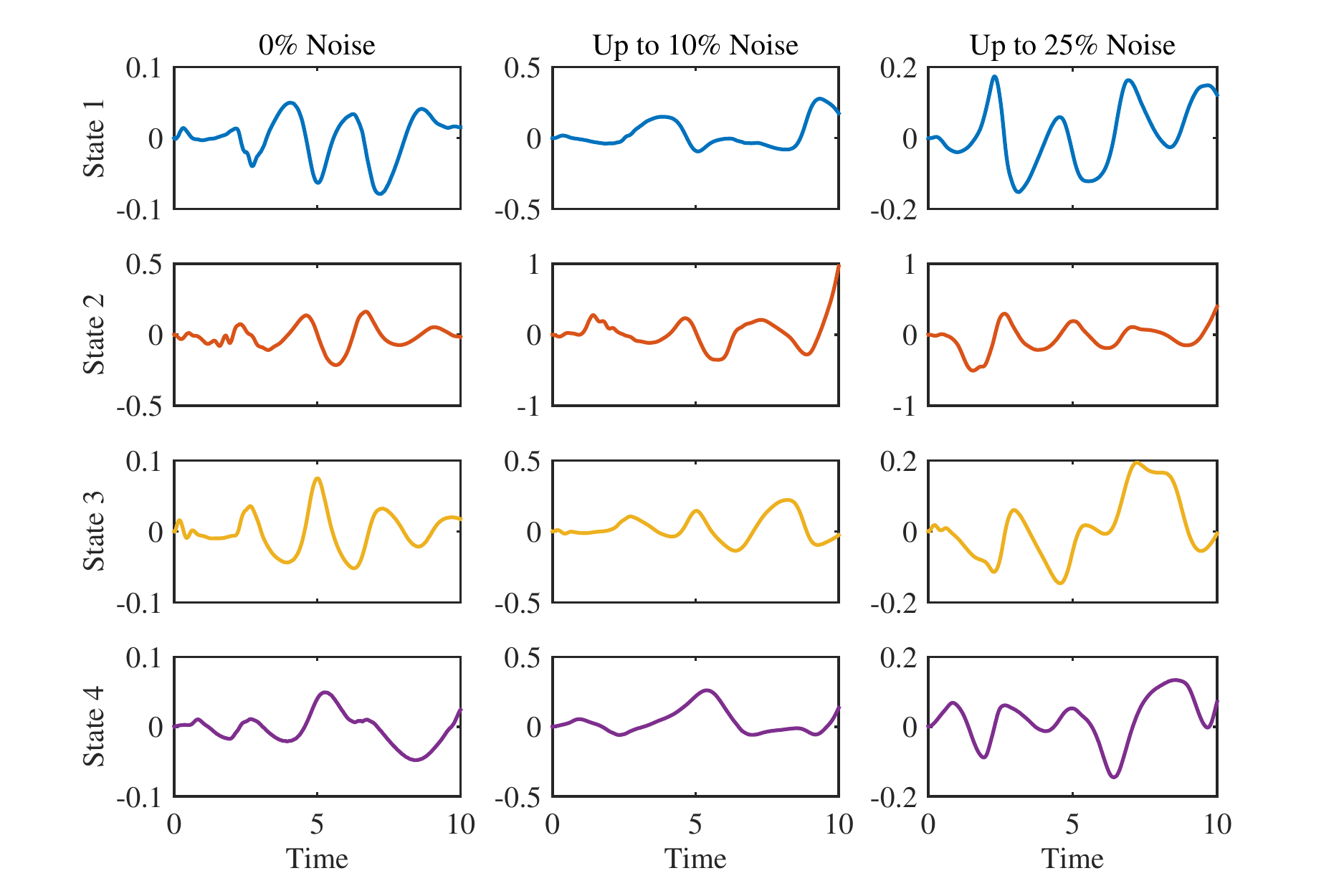}
\caption{{\bf Relative error between spline solutions and state solutions.} \\ Each row shows the relative error for a single state for all three simulation studies. The relative error between the spline solution and state solution is calculated as $\frac{s-y}{y}$ over the time interval $[0,10]$ with $s$ denoting the optimal spline approximation and $y$ denoting the numerical solution of the optimal Lotka-Volterra system.}
\label{fig:Fig4}
\end{figure}

By plotting the relative error between the spline solutions and the state solutions in Figure~\ref{fig:Fig4}, all three studies additionally illustrated that the optimal spline functions also proved to be a good approximation of the state solutions found using the optimal model parameters.

\subsection{Interactions within the Intestinal Microbiota}\label{sub:microbiota}
The generalized Lotka-Volterra formalism was recently used to explore the impact of the intestinal microbiota on the development of antibiotic-induced {\it Clostridium difficile} colitis~\cite{Stein2013}. This disease occurs in patients who have been treated with antibiotics and is characterized by a marked shift of the intestinal microbiota towards a state dominated by the pathogen {\it Clostridium difficile}. In many cases health can only be restored through a fecal transplant, a procedure which restores the diversity of the microbiota. The mechanisms through which disease occurs, and through which the normal gut microbiota can prevent the over-growth of {\it C. difficile} are still not well understood. 

In Stein et al.~\cite{Stein2013} the authors relied on a mouse model of {\it C. difficile} colitis to attempt to address these questions. They tracked the microbiota of mice across time and used the resulting data to estimate the parameters of a Lotka-Volterra model. Based on the resulting model they were able to provide new testable hypotheses about the factors that promote the overgrowth of {\it C. difficile} following a course of clindamycin. Here we used the same data and model and added the assumption of interaction sparsity to test our parameter estimation procedure and compare to the originally published results. 

We focused on a subset of the Stein et al. data, specifically data originating from three mice who had not been subjected to any antibiotic interventions. The exact details, including how we setup our problem, can be found in Appendix~\ref{apx:stein}.

\begin{figure}[H]
\includegraphics[width=\textwidth]{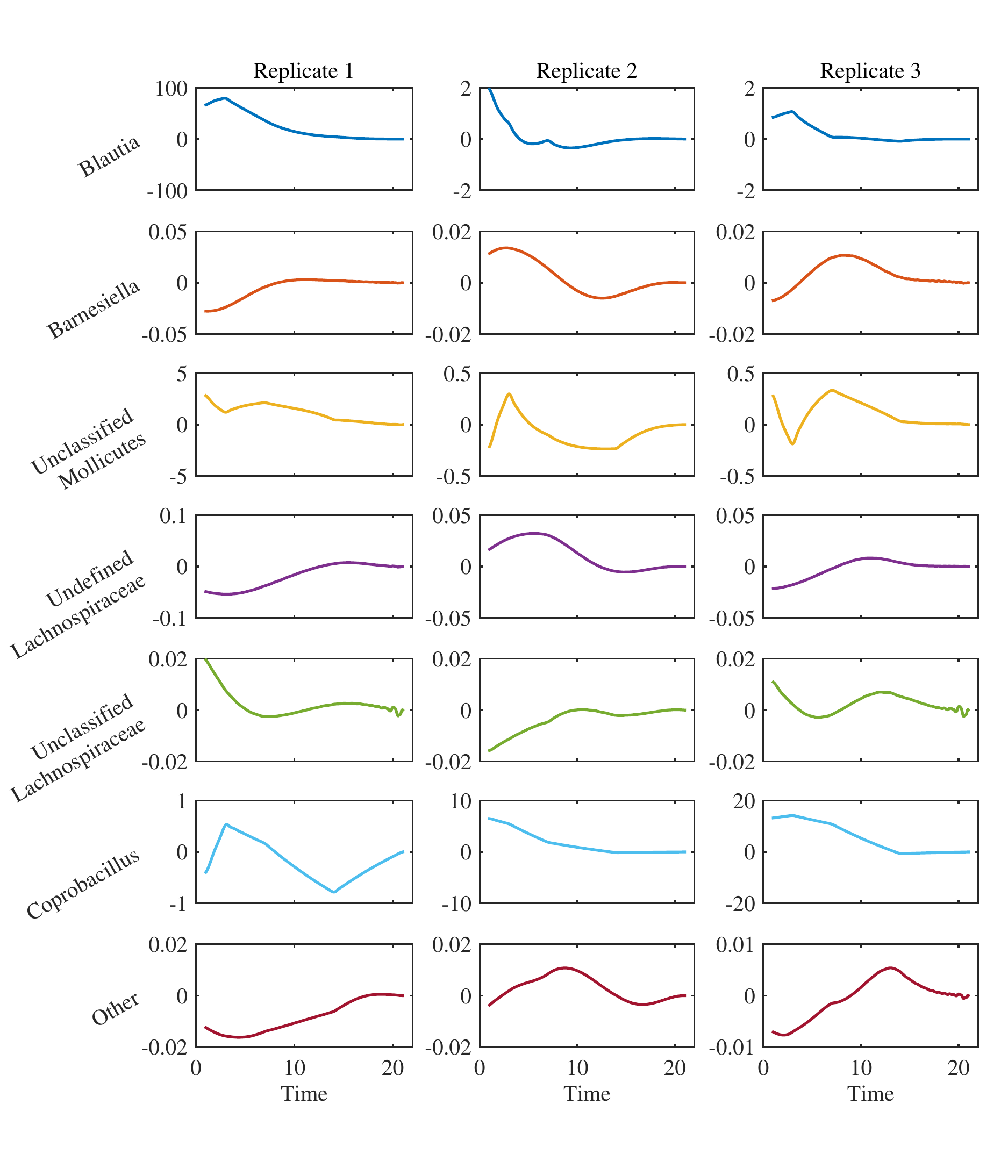}
\caption{{\bf Relative error between spline solutions and state solutions.} \\ Each row shows the relative error for a single OTU for all three replicate experiments. The relative error between the spline solution and state solution is calculated as $\frac{s-y}{y}$ over the time interval $[1,21]$ with $s$ denoting the optimal spline approximation and $y$ denoting the numerical solution of the optimal Lotka-Volterra system. Time is measured in days.}
\label{fig:Fig5}
\end{figure}

In our simulation, the spline solutions remained good approximations to the state solutions as demonstrated in Figure~\ref{fig:Fig5}. The relative errors for the spline solutions for the \textit{Blautia} and \textit{Coprobacillus} OTUs were larger than the relative errors for the other five OTUs, but this was due to both the significantly smaller magnitude of the data for these OTUs and the magnitude of the weights for the data relative to the other OTUs. Among the three replicates for a single OTU, variations in the magnitude of the relative errors, e.g., in \textit{Blautia}, \textit{Unclassified Mollicutes}, and \textit{Coprobacillus}, were explained by noticeable variations in the magnitude of the data across replicates. 

\begin{figure}[H]
\includegraphics[width=\textwidth]{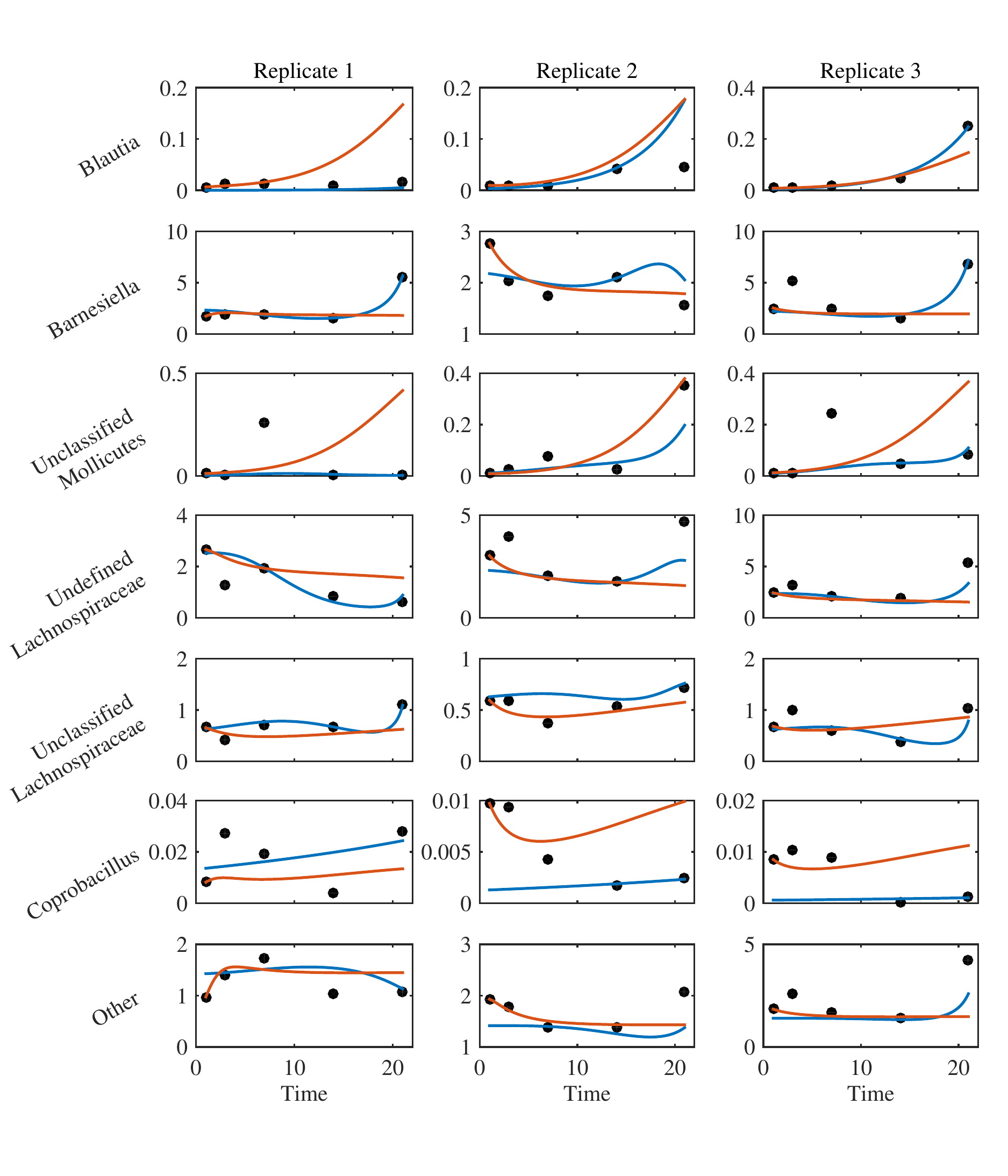}
\caption{{\bf Comparison of state solutions.} \\ Each row compares the state solutions for a single OTU for all three replicate experiments using our parameter estimation results and those found in Stein et al.~\cite{Stein2013}. The black dots indicate the provided experimental data. The blue curves mark the state solutions found using our optimal Lotka-Volterra system, and the red curves denote the state solutions found using the optimal Lotka-Volterra system published by Stein et al. Time is measured in days, and abundance is measured in $10^{11}$ DNA copies per cubic centimeter.}
\label{fig:Fig6}
\end{figure}

As in the simulation studies, we also assess our method's ability to recover the data. The relative error between the state solutions using the optimal model parameters and the data for all three mice, given by $e_{r} = \frac{1}{m}\norm[1]{\frac{\bfm(\bfy)-\bfd}{\bfd}}$ with $m$ being the number of elements in $\bfd$ and the division being element-wise, was $e_{r} \approx 0.4594$.  The relative error for the model published in~\cite{Stein2013} was $e_{r} \approx 3.6790$, indicating that our method more accurately captured the dynamics of the data. This fact was further confirmed by a visual comparison of the state solutions to the data in Figure~\ref{fig:Fig6}.

\begin{figure}[H]
\includegraphics[width=\textwidth]{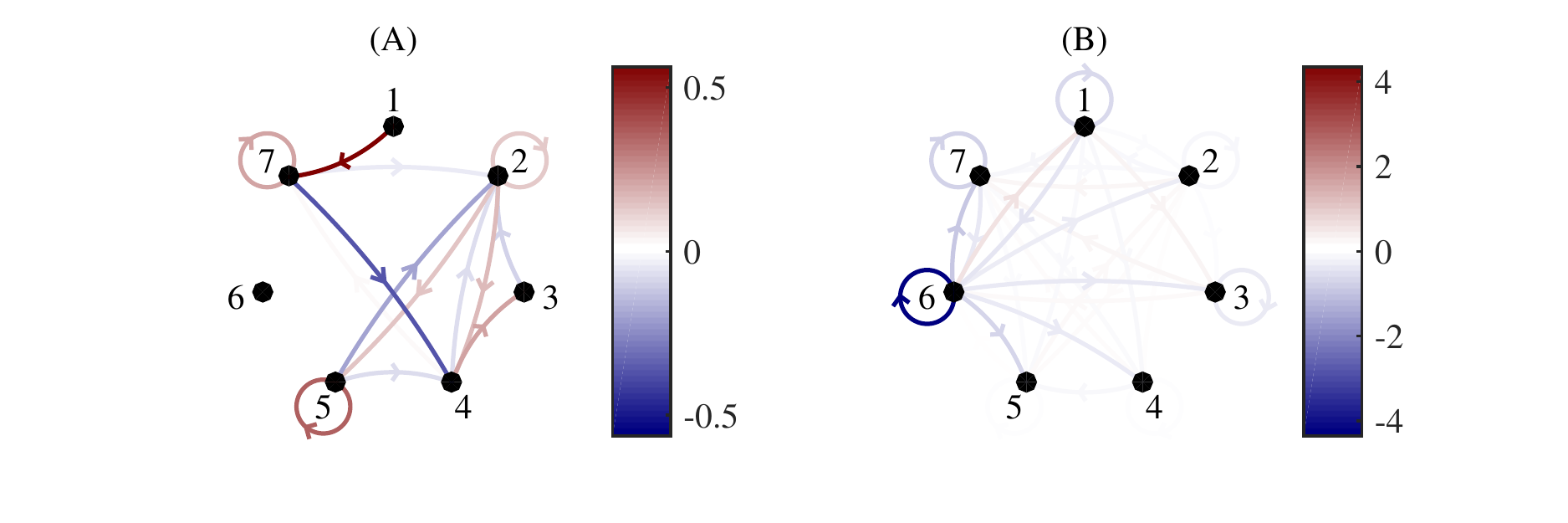}
\caption{{\bf Comparison of interaction matrices.} \\ This figure compares (A) the interaction matrix from our optimal Lotka-Volterra system to (B) a subset of the interaction matrix published by Stein et al.~\cite{Stein2013}. In the graphs, entry $a_{ij}$ in the interaction matrix is given as a directed edge from node $j$ to node $i$. The value of the entry $a_{ij}$ is given by the color of the edge. OTUs: 1-\textit{Blautia}, 2-\textit{Barnesiella}, 3-\textit{Unclassified Mollicutes}, 4-\textit{Undefined Lachnospiraceae}, 5-\textit{Unclassified Lachnospiraceae}, 6-\textit{Coprobacillus}, and 7-\textit{Other}.}
\label{fig:Fig7}
\end{figure}

The biological implication of the disagreement between our results and those originally published on the same data became apparent when examining graph representations of the Lotka-Volterra interaction matrices in Figure~\ref{fig:Fig7} and considering a recent paper from the same group providing experimental evidence for the role of the gut microbiota in the prevention of \textit{C. difficile} infection~\cite{Buffie2015}. Stein et al. originally concluded, on the basis of Lotka-Volterra modeling, that members of the \textit{Coprobacillus} genus inhibit the growth of other members of the gut microbiome, which implied \textit{Coprobacillus} is a stabilizing factor within the gut microbiome. In our own analysis of their data, we did not identify any strong interactions between the \textit{Coprobacillus} OTU and other organisms. Instead we observed inhibitory interactions of members of the \textit{Lachnospiraceae} family with other gut microbes, which suggested that members of the \textit{Undefined Lachnospiraceae} and \textit{Unclassified Lachnospiraceae} groups are the more likely players involved in preventing \textit{C. difficile} colonization. Buffie et al. confirmed this experimentally~\cite{Buffie2015}. In their paper they showed that the Lachnospiraceae species \textit{Clostridium scindens} can provide resistance to \textit{C. difficile} colonization in a mouse model of \textit{C. difficile} enterocolytis.  

\section{Discussion} \label{sec:dis}
\subsection*{Method Analysis}
In the Methods section we derive our approach given by~\eqref{eq:pe4} from the typical parameter estimation problem given by~\eqref{eq:pe1}. In our approach, we do not find a solution to the original problem statement, but we solve a ``nearby" problem instead with the idea that~\eqref{eq:pe4} is numerically more robust. It is feasible to criticize that the optimization problem being solved is just an approximation to the solution of~\eqref{eq:pe1}, but even in cases where the solution to~\eqref{eq:pe4} is not sufficiently accurate, this method can be used to efficiently precompute approximations for $\hat\bfp$ and $\hat\bfy_0$, which can then be used as initial guesses for single or multiple shooting methods.

One step in constructing our ``nearby" problem is replacing the state variable $\bfy$ in the model with an approximation $\bfs$. In our case, we use cubic spline functions for $\bfs$, and for $\bfs =  [s_1,\ldots, s_n]\t$ the approximation error for each $s_j \in \calS^3_\bftau([a,b])$, $j = 1,\ldots,n$, is bounded using the theorem below. 
\begin{thm}[\cite{Powell1981}]
	Let $m$ be a positive integer. For every $y \in \calC^m([a,b])$ and for every integer $j\in \{1, \ldots, \min(m,4)\}$, the least maximum error satisfies the condition
	\begin{equation*}
	\min_{s \in \calS^3_\bftau([a,b])} \norm[\infty]{y - s} \leq \frac{4!}{(4-j)!} \frac{1}{2^j}h^j \norm[\infty]{y^{(j)}},
	\end{equation*}
	where $h = \max\set{\tau_{i+1}-\tau_{i}: i = 0,\ldots,k-1}$.
\end{thm}
\noindent Since $y_j \in \calC^1([a,b])$ for $j=1,\ldots,n$ the bound on the approximation error for each $s_j \in \calS^3_\bftau([a,b])$, $j = 1,\ldots,n$, simplifies to 
\begin{equation*}\label{eq:thm}
\min_{s \in \calS^3_\bftau([a,b])} \norm[\infty]{y - s} \leq 2 h \norm[\infty]{y'}.
\end{equation*}

Also, while the ``nearby problem" is numerically more robust than the problem given in~\eqref{eq:pe1}, the dimension of the optimization problem increases. This can adversely affect the speed of the optimization step, but is also counteracted by improvements in computational efficiency elsewhere. One example is that optimization steps in~\eqref{eq:pe4} never require the calculation of the solution of the ODE model. Eliminating the need for the solution of the initial value problem removes a computationally intensive step in each optimization iteration and replaces  the step with the analytic evaluation of the spline vector $\bfs$ and its time derivative $\bfs'$. 

\begin{figure}[h]
\includegraphics[width=\textwidth]{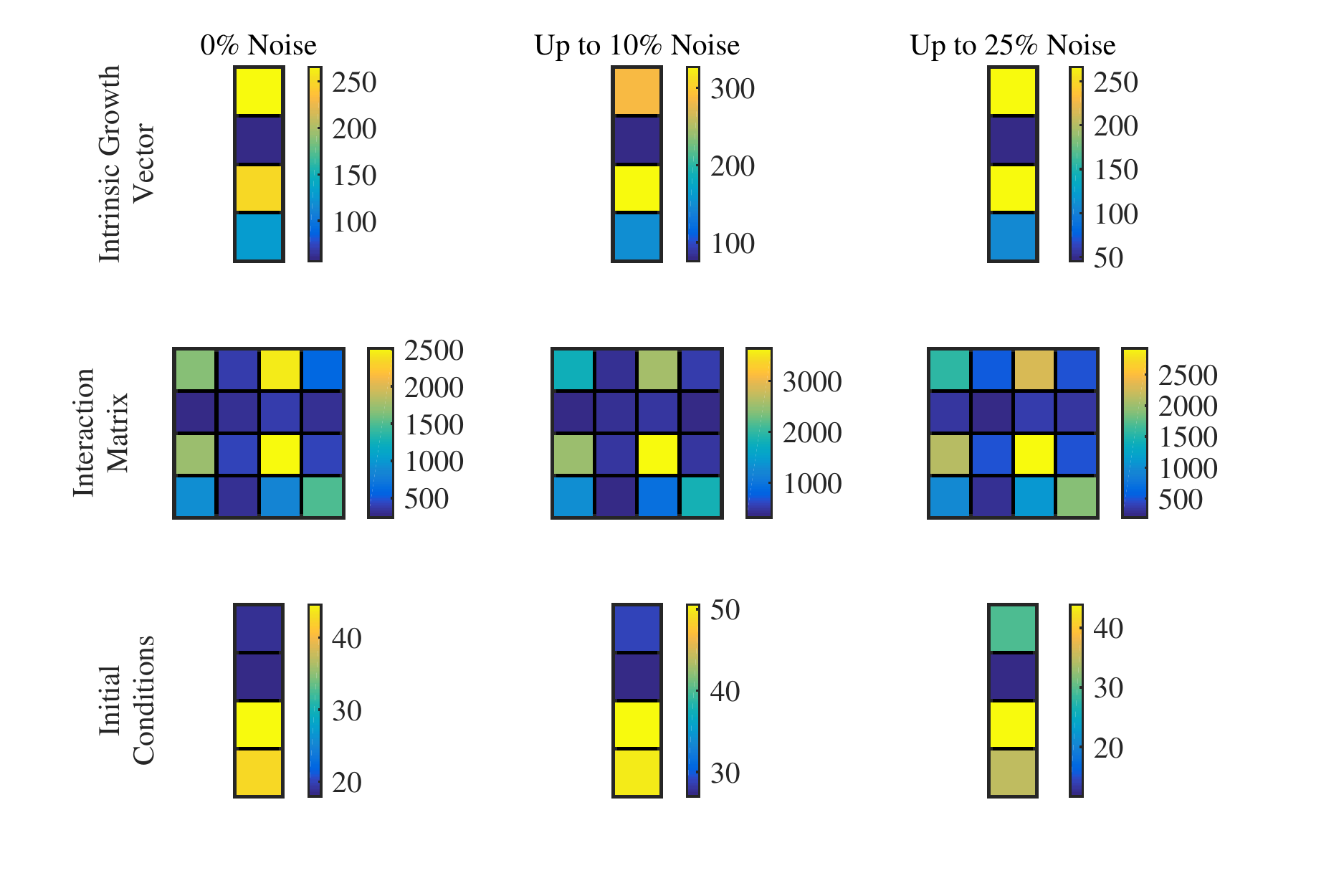}
\caption{{\bf Cumulative local sensitivities of optimized model parameters and initial conditions.} \\ The images display the cumulative local sensitivities for every optimal model parameter and initial condition for all three simulation studies.}
\label{fig:Fig8}
\end{figure}

\subsection*{Simulation Studies} \label{sub:sim}
As an unknown sparsity pattern was an assumption in our model for these simulation studies, our method's ability to capture the true sparsity pattern is relevant. Due to numerical and computational limitations, it is unlikely for any model parameter in the optimal set to be identically 0, so we instead consider any model parameters with a magnitude below a certain threshold, $10^{-3}$ in this case, to essentially be 0. The sparsity structure is perfectly preserved in the first simulation study, 87.5 percent recovered in the second, and 62.5 percent recovered in the third. This is what we would expect as the increase in data noise over the three simulation studies should have an increasingly significant effect on the accuracy of the model parameters that our method returns. 

A model's local sensitivity to the optimal model parameters and initial conditions, which is the solution of the initial value problem
\begin{equation*}\label{eq:sens}
\frac{\rm d}{{\rm d}t}\left(\begin{bmatrix} 
\bfy \\[1ex] 
\frac{\partial \bfy}{\partial \bfp} \\[1ex] 
\frac{\partial \bfy}{\partial \bfy_0} 
\end{bmatrix}\right) = 
\begin{bmatrix} 
\bff(t, \bfy, \bfp) \\[1ex]
\frac{\partial}{\partial \bfp}\bff(t, \bfy, \bfp) \\[1ex] 
\frac{\partial}{\partial \bfy_0}\bff(t, \bfy, \bfp) 
\end{bmatrix}, \quad
\begin{bmatrix}
\bfy(a) \\[1ex]
\frac{\partial}{\partial \bfp}\bfy(a) \\[1ex]
\frac{\partial}{\partial \bfy_0}\bfy(a) 
\end{bmatrix} = 
\begin{bmatrix} 
\bfy_0 \\[1ex]
\bfzero \\[1ex]
\vec{\bfI_n}
\end{bmatrix}
\end{equation*}
over the interval $[a,b]$, is also often of interest. To have some idea of the cumulative effect of a single parameter or initial condition on the entire $n$-state Lotka-Volterra system we calculate $S_{p_i}$ for $i = 1,\ldots,n^2+n$  and $S_{y_{0,i}}$ for $i = 1,\ldots,n$ with
\begin{equation*}
S_{p_i} = \sum_{j=1}^n \int_a^b \abs{\frac{\partial y_j(t)}{\partial p_i}} dt \quad \mbox{and} \quad S_{y_{0,i}} = \sum_{j=1}^n \int_a^b \abs{\frac{\partial y_j(t)}{\partial y_{0,i}}} dt, 
\end{equation*}
respectively, and provide the results in Figure~\ref{fig:Fig8}.

The cumulative sensitivities remain consistent across the three simulation studies, which is to be expected given the consistency across the simulation studies in the optimal model parameters and initial conditions themselves. 

\subsection*{Interactions within the Intestinal Microbiota}
\begin{figure}[H]
\includegraphics[width=\textwidth]{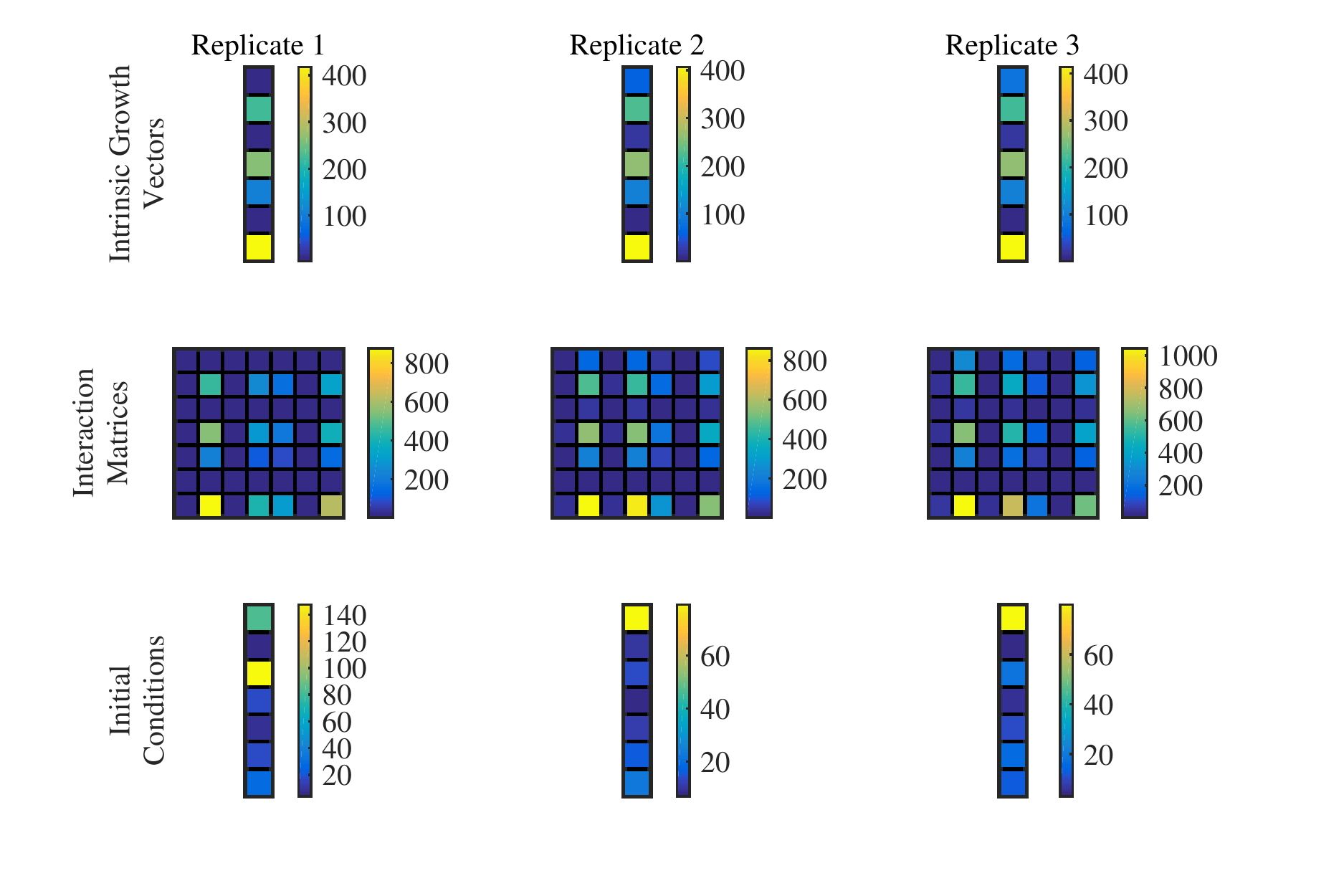}
\caption{{\bf Cumulative local sensitivities of optimized model parameters and initial conditions.} \\ The images display the cumulative local sensitivities for every optimal model parameter and initial condition for all three replicate experiments.}
\label{fig:Fig9}
\end{figure}

As with the simulation studies, we can calculate the cumulative local sensitivities with respect to the optimal model parameters and initial conditions. Note that because of how the parameter estimation problem is setup (see Appendix~\ref{apx:stein}), the resulting model parameters for each of the three replicates are the same, but the optimal initial conditions differ. This means the local sensitivities and hence the cumulative sensitivities can vary by replicate, yet the results in Figure~\ref{fig:Fig9} display consistency in the cumulative sensitivities across the replicates.

Biologically speaking, our method demonstrates the ability to closely model real biological data, but we would like to note that the resulting model is still unable to provide a full forward prediction of the intestinal microbiome's state. The Lotka-Volterra system appears to be unstable leading to the uncontrolled growth, or the disappearance of certain taxa, phenomena not commonly observed in real data. In part this is due to the limitations of the Lotka-Volterra system and the simplifying assumptions made when choosing it as a model. Insufficient data both in terms of the relatively small number of samples and, more importantly, in terms of the sparse sampling rate also play a role. Unfortunately, these limitations are inherent as computational costs can require model simplifications, experimental costs can limit the number of feasibly obtained samples, and the specific microbiome being sampled can potentially limit the sampling frequency. 

Despite these limitations, modeling approaches such as the few described here are still extremely useful in understanding host-associated microbial communities. In particular, we showed that we could identify interactions and their direction between members of the gut microbiome that were later confirmed by in vivo experiments. The mathematical framework we described here correctly identified the \textit{Lachnospiraceae} family as playing a stabilizing role in the gut microbiome, which contrasts with the suggestion that the \textit{Coprobacillus} OTU plays this role as previously predicted, and while the full dynamics of the system cannot yet be predicted, knowledge of the interactions and their direction can sufficiently guide further biological experimentation. Therefore, we suggest that computational approaches such as ours effectively combined with experimental approaches can help elucidate the role of host-associated microbes in health and disease. 

\section*{Acknowledgments}
This work was supported by the National Institute Of General Medical Sciences of the National Institutes of Health under Award Number R21GM107683. The authors would like to thank Robert Torrence from Virginia Tech and Hector Bravo and Senthil Muthiah from the University of Maryland, College Park for providing comments and feedback.

\bibliographystyle{siam} 
\bibliography{refs}

\newpage

\appendix

\section{Cubic Spline Construction}
\label{S1}
Using gradient-based optimization methods for solving~\eqref{eq:pe4} requires the derivatives of the spline $s$ with respect to the time and the coefficients $\bfq$. In particular we need to compute $s$, $s_\bfq$, $s'$, and $s'_\bfq$. Here, we provide the required derivations for the cubic spline function $s$ and its derivatives.

\subsection{Spline Definition} \label{apx:splines}
Let data points $a = t_0<\cdots< t_n = b$ and any corresponding real numbers $q_j$, $j = 0,\ldots,n$ been given.\\

Let $s:[a,b] \to \bbR$ be a function with the interpolation properties
\begin{equation}
	\label{eq:interpolation} s(t_j) = q_j \quad \mbox{for} \quad j = 0,\ldots,n 
\end{equation}
that is equal to a cubic polynomial with coefficients $a_j$, $b_j$, $c_j$, and $d_j$ on each interval $[t_j, t_{j+1}]$, i.e., 
\begin{equation}
	\label{eq:pp} s(t) \big|_{t \in [t_j, t_{j+1}]} = s_j(t) = a_j (t-t_j)^3 + b_j (t-t_j)^2 + c_j (t-t_j) + d_j 
\end{equation}
for $ t_j \leq t < t_{j+1}$, $j = 0,\ldots,n-1$, and is twice differentiable, i.e.,
\begin{equation}
	\label{eq:continuous} s_j'(t_{j+1}) = s_{j+1}'(t_{j+1}) \quad \mbox{and} \quad s_j''(t_{j+1}) = s_{j+1}''(t_{j+1}) 
\end{equation}
for $j = 0,\ldots,n-2$. We refer to $s$ as a cubic spline~\cite{deBoor2001}.\\

Equations~\eqref{eq:interpolation},~\eqref{eq:pp} and~\eqref{eq:continuous} provide $4n-2$ conditions for the $4n$ coefficients $a_j$, $b_j$, $c_j$, and $d_j$,  $j = 0,\ldots,n-1$. A spline $s$ is now uniquely determined by choosing appropriate boundary conditions. Here, we choose not-a-knot boundary conditions, i.e., $s_0'''(t_1) = s_1'''(t_1)$ and $s_{n-2}'''(t_{n-1}) = s_{n-1}'''(t_{n-1})$. 
For efficient calculations of the coefficients $a_j$, $b_j$, $c_j$, and $d_j$, we define moments
\begin{equation*}\label{eq:moments}
	m_j = s''(t_j)
\end{equation*}
for $j = 0,\ldots,n.$ Note that the second derivative of $s$ is linear in each interval $[t_j, t_{j+1}]$, and for $t \in [t_j, t_{j+1}]$, the moments $m_j$ have the expression
\begin{equation*}\label{eq:ddspline}
	s_j''(t) = m_j \frac{t_{j+1} - t }{h_{j+1}} + m_{j+1} \frac{t - t_j }{h_{j+1}}
\end{equation*}
with $h_{j+1} = t_{j+1} - t_{j}$ for $j = 0,\ldots,n-1$. Integrating and using equation~\eqref{eq:interpolation} we get 
\begin{equation*}\label{eq:dspline}
	s_j'(t) = -m_j \frac{(t_{j+1} - t)^2}{2h_{j+1}} + m_{j+1} \frac{(t - t_j)^2}{2h_{j+1}} + \frac{q_{j+1} - q_j}{h_{j+1}} - \frac{h_{j+1}}{6}(m_{j+1} - m_j)
\end{equation*}
\begin{multline*}\label{eq:spline}
	s_j(t) = m_j \frac{(t_{j+1} - t)^3}{6h_{j+1}} + m_{j+1} \frac{(t - t_j)^3 }{6h_{j+1}} \\ + \left( \frac{q_{j+1} - q_j}{h_{j+1}} - \frac{h_{j+1}}{6}(m_{j+1} - m_j) \right) (t - t_j) + q_j - m_j \frac{h_{j+1}^2}{6}. 
\end{multline*}
With the condition $s_j'(t_{j+1}) = s_{j+1}'(t_{j+1})$ for $j = 1,\ldots,n-1$ we get the equation 
\begin{equation*}\label{eq:meq1}
	\frac{h_j}{6} m_{j-1} + \frac{h_j + h_{j+1}}{3} m_j + \frac{h_{j+1}}{6} m_{j+1} = \frac{q_{j+1} - q_j}{h_{j+1}} - \frac{q_j - q_{j-1}}{h_j},
\end{equation*}
which can be rewritten as
\begin{equation*}\label{eq:meq2}
	(1 - \lambda_j) m_{j-1} + 2 m_j + \lambda_j m_{j+1} = \frac{6}{h_j + h_{j+1}} \left(\frac{q_{j+1} - q_j}{h_{j+1}} - \frac{q_j - q_{j-1}}{h_j}\right) 
\end{equation*}
with $\lambda_j = \frac{h_{j+1}}{h_j + h_{j+1}}$ for $j = 1,\ldots,n-1$.\\

Now the moments $\bfm = [m_0,\ldots,m_n]\t$ can be calculated by solving the linear system 
\begin{equation*}\label{eq:momentsys}
	\bfA \bfm = \bfbeta. 
\end{equation*}
Here, $\bfbeta = [\beta_0,\ldots,\beta_n]\t$ with $\beta_j = \frac{6}{h_j + h_{j+1}} \left(\frac{q_{j+1} - q_j}{h_{j+1}} - \frac{q_j - q_{j-1}}{h_j}\right)$ for $j = 1,\ldots,n-1$ and 
\begin{equation*}
	\bfA = 
	\begin{bmatrix*}
		\bfA_0\\
		\tilde{\bfA}\\
		\bfA_n 
	\end{bmatrix*}
	\in \bbR^{(n+1) \times (n+1)} 
\end{equation*}
with 
\begin{equation*}
	\tilde{\bfA} = 
	\begin{bmatrix*}
		1-\lambda_1 & 2 & \lambda_1 & & & & \\
		& \ddots & \ddots & \ddots & & & \\
		& & \ddots & \ddots & \ddots & & \\
		& & & \ddots & \ddots & \ddots & \\
		& & & & 1-\lambda_{n-1} & 2 & \lambda_{n-1} \\
	\end{bmatrix*}.
\end{equation*}
The entries $\bfA_0, \bfA_n$, $\beta_0$, and $\beta_n$ are specified by the boundary conditions. For the not-a-knot conditions,
\begin{align*}
\bfA_0 &= \left[ -\frac{1}{h_1}, \frac{1}{h_1} + \frac{1}{h_2}, -\frac{1}{h_2}, 0, \ldots, 0 \right], \\
\bfA_n &= \left[ 0, \ldots, 0, -\frac{1}{h_{n-1}}, \frac{1}{h_{n-1}} + \frac{1}{h_n}, -\frac{1}{h_n} \right], \\
\beta_0 &= \beta_n = 0.
\end{align*} 
With $\bfm$ given, coefficients $a_j$, $b_j$, $c_j$, and $d_j$ are determined by
\begin{align}\label{eq:coef}
	\begin{split}
	d_j &= q_j \\
	c_j &= \frac{q_{j+1} - q_j}{h_{j+1}} - \frac{(2m_j + m_{j+1})h_{j+1}}{6} \\
	b_j &= \frac{m_j}{2} \\
	a_j &= \frac{m_{j+1} - m_j}{6 h_{j+1}} 
	\end{split}
\end{align}
for $j = 0,\ldots,n-1$.

\subsection{Derivatives of the Spline Function}\label{apx:splinesD}
We are interested in calculating $s_\bfq$, where $\bfq = [q_0,\ldots,q_n]\t$. With equation~\eqref{eq:pp} we get 
\begin{align*}\label{eq:dsdq}
	s_\bfq(t) \big|_{t \in [t_j,t_{j+1}]} &= \frac{\d s_j}{\d \bfq} (t)\\
	&= \frac{\d}{\d \bfq} \left(a_j (t-t_j)^3 + b_j (t-t_j)^2 + c_j (t-t_j) + d_j\right) \\
	&= (t-t_j)^3 \frac{\d a_j}{\d \bfq}  + (t-t_j)^2 \frac{\d b_j}{\d \bfq}  + (t-t_j) \frac{\d c_j}{\d \bfq} + \frac{\d  d_j}{\d \bfq}. 
\end{align*}
First, the coefficients $a_j$, $b_j$, $c_j$, and $d_j$ depend on the moments $\bfm$. The derivative $\frac{\d \bfm}{\d \bfq} = \frac{\d}{\d \bfq}(\bfA^{-1} \bfbeta) = \bfA^{-1} \frac{\d \bfbeta}{\d \bfq} = \bfA^{-1}\bfB$ with 
\begin{equation*}
	\bfB = \frac{\d \bfbeta}{\d \bfq} = 
	\begin{bmatrix*}
		0 & 0 & 0 & & & \\
		-\mu_1 \lambda_1 & \mu_1 & -\mu_1 (1 - \lambda_1) & & & \\
		& -\mu_2 \lambda_2 & \mu_2 & -\mu_2 (1 - \lambda_2) & & \\
		& & \ldots & \ldots & \ldots & \\
		& & & -\mu_{n-1} \lambda_{n-1} & \mu_{n-1} & -\mu_{n-1} (1 - \lambda_{n-1}) \\
		& & & 0 & 0 & 0 \\
	\end{bmatrix*}
\end{equation*}
for the not-a-knot boundary conditions and for $\mu_j = -\frac{6}{h_j h_{j+1}}$, $j = 1,\ldots,n-1$. We can calculate $\frac{\d \bfm}{\d \bfq}$ by solving the linear systems $\bfA\bfM = \bfB$ where $\bfM = \frac{\d \bfm}{\d \bfq}$.

Finally, the derivatives of the coefficients with respect to $\bfq$, i.e., $\frac{\d \bfa}{\d \bfq}$,  $\frac{\d \bfb}{\d \bfq}$, $\frac{\d \bfc}{\d \bfq}$, and $\frac{\d \bfd}{\d \bfq}$ with $\bfa = [a_0,\ldots,a_{n-1}]\t, \bfb = [b_0,\ldots,b_{n-1}]\t, \bfc = [c_0,\ldots,c_{n-1}]\t$ and $\bfd = [d_0,\ldots,d_{n-1}]\t$, can be computed using~\eqref{eq:coef} and are given by
\begin{align*}\label{eq:dcoefs}
		\frac{\d \bfd}{\d \bfq} &= \bfE_n \\
		\frac{\d \bfc}{\d \bfq} &= \bfH_{-1} \odot (\bfE_0 - \bfE_n) - \frac{1}{6} \bfH_1 \odot (2 \bfE_0 + \bfE_n) \bfM \\
		\frac{\d \bfb}{\d \bfq} &= \frac{1}{2}\bfE_n \bfM \\
		\frac{\d \bfa}{\d \bfq} &= \frac{1}{6} \bfH_{-1} \odot (\bfE_0 - \bfE_n) \bfM,
\end{align*}
where $\bfE_0 = [\bfzero, \,\bfI_n]$, $ \bfE_n = [\bfI_n, \bfzero]$, $\bfI_n$ is the $n\times n$ identity matrix, $\bfH_1 = [h_1,\ldots,h_n]\t \otimes [1,\ldots,1]$, and $\bfH_{-1} = [1/h_1,\ldots,1/h_n]\t \otimes [1,\ldots,1]$. The symbols $\otimes$ and $\odot$ denote the Kronecker product the Hadamard product, respectively.

The further required derivatives $s'$ and $s'_\bfq$ can now be easily computed by
\begin{equation*}\label{eq:dsdt}
	s'(t) \big|_{t \in [t_j, t_{j+1}]} = 3 a_j (t-t_j)^2 + 2 b_j (t-t_j) + c_j 
\end{equation*}
and
\begin{equation*}\label{eq:dsdqdt}
	s'_\bfq(t) \big|_{t \in [t_j,t_{j+1}]} = 3 (t-t_j)^2 \frac{\d a_j}{\d \bfq} + 2 (t-t_j) \frac{\d b_j}{\d \bfq} + \frac{\d c_j}{\d \bfq}. 
\end{equation*}

\section{Computational Details} \label{apx:comp}
\label{S2}
To run continuous shooting with a gradient-based optimization method on a problem that uses a Lotka-Volterra model with assumptions on sparsity, we require specific derivatives of the model and a mathematical formulation of the sparsity assumption. Here, we include these details and also our step-by-step approaches to both the simulation studies and the intestinal microbiota example.

\subsection{Numerics for Lotka-Volterra Models and Sparsity Constraints}
Using the Lotka-Volterra system~\eqref{eq:lv} as our model, we assume the parameters $\bfp = [\bfb; \vec{\bfA}]$ are considered unknown. Let us also replace the model state variable $\bfy$ with its approximation $\bfs$. The derivatives of $\bff$ with respect to $\bfs$ and $\bfp$ are given by
\begin{align*}\label{eq:lvderivs}
	\bff_\bfs &= \diag{\bfb} + \diag{\bfs}\bfA + \diag{\bfA\bfs}\\
	\bff_\bfp &= \begin{bmatrix*} \diag{\bfs}, \bfs\t \otimes \diag{\bfs} \end{bmatrix*}
\end{align*}
with $\otimes$ denoting the Kronecker product. 

To include a sparsity constraint on the interaction matrix $\bfA$ in~\eqref{eq:pe4}, the constraint term $\alpha\calD(\bfc(\bfp,\bfs(t;\bfq))$ becomes $\alpha\norm[1]{\vec{\bfA}}$. In this case $\calD$ is the one-norm, and $\bfc$ is the function that maps $\bfp$ to a vector of the parameters in $\bfA$. Note that this constraint is not differentiable everywhere, but one way to overcome this is by approximating the 1-norm using a smooth function. The approximation we use is
\begin{equation}\label{eq:sparsity}
\norm[1]{\vec{\bfA}} \approx \sum_{i=n+1}^{n^2+n} H_\eps(p_i)
\end{equation}
where $H_\eps$ is the Huber function defined by
\begin{equation*}
H_\eps(x) = 
\begin{dcases}
x - \frac{\eps}{2}, & \abs x  \geq \eps \\
\frac{x^2}{2\eps}, & \abs x < \eps.
\end{dcases}
\end{equation*}
The idea here is that the function $\norm[1]{\vec{\bfA}}$ is approximated by a smooth quadratic curve near its corners with ``near" being defined by the choice of $\eps$. Another important note regarding this modification to the objective function is that the effect on the numerics of the method. The data-fitting and model-fitting contributions to the function, gradient, and Hessian terms can be calculated as before. The contributions of the sparsity constraint term, however, require the first and second derivatives with respect to $\bfp$ and $\bfq$ of the approximation given in~\eqref{eq:sparsity}.

\subsection{Simulation Study Details}\label{apx:simu}
Given the Lotka-Volterra system with parameters and initial conditions 
\begin{equation*}
\bfb = \begin{bmatrix*}[r] 2 \\ 1\\ 0 \\ -3 \end{bmatrix*}, \quad \bfA = \begin{bmatrix*}[r] 0 & -0.6 & 0 & -0.2 \\ 0.6 & 0 & -0.6 & -0.2 \\ 0 & 0.6 & 0 & -0.2 \\ 0.2 & 0.2 & 0.2 & 0 \end{bmatrix*}, \quad \mbox{and} \quad \bfy_0 = \begin{bmatrix*}[r] 5 \\ 4 \\ 3 \\ 2 \end{bmatrix*},
\end{equation*}
we numerically solve the initial value problem and let $\bfy_{\rm true}$ denote this forward solution. We then collect the values of $\bfy_{\rm true}$ at the times given by the uniform discretization $0 = t_1<\ldots<t_{20} = 10$ and perturb them to generate the data $\bfd = \vec{\bfD} = \vec{[\bfd_1,\ldots,\bfd_{20}]}$. Here, $\bfd_j = (1 + \bfeps_j)\bfy_{{\rm true}, j}$ for $j = 1,\ldots,20$ with $\bfeps_j$ representing a scaled vector of independent and identically Beta distributed noise, i.e., $\bfeps_j \sim \gamma \cdot({\rm Beta}(2,2)-1/2)$. We conduct three studies, each with different noise level scales (Study 1: $\gamma = 0$; Study 2: $\gamma = 0.1$; Study 3: $\gamma = 0.25$).

In the objective function, the projection $\bfm$ is the identity projection for all three studies, but the weight matrix $\bfW$ is different for each study because it depends on the data and is taken to be $\bfW = \bfI_{20} \otimes \diag{[w_1,\dots,w_4]\t}$ where $\bfI_{20}$ is the $20\times20$ identity matrix and $w_i = 10/\sigma(\bfD_i)$, $i = 1,\ldots,4$, is the linearly scaled weighting of the inverse standard deviation of state $i$'s time-series data. Additionally, we use the standard constraint term in the objective function with a sparsity constraint on the interaction matrix $\bfA$, which was defined earlier. For each study, we separately sample 1,000 $(\lambda, \alpha)$-pairs from the square $[1, 100] \times [0.01, 1]$ and choose a pair using leave-one-out cross-validation. The $(\lambda, \alpha)$-pairs are approximately $(1.1416, 0.01261)$, $(5.5098, 0.04584)$, $(23.8228, 0.87599)$ for studies 1, 2, and 3, respectively.

For each study, we then separately sample the parameter space 10,000 times using a Latin hypercube sampling and perform local optimizations using the Gauss-Newton method with each sample serving as an initial parameter set. The global minimizer is the local minimizer that most minimizes the objective function.

\subsection{Intestinal Microbiota Details}\label{apx:stein}
The data collected consists of the abundance levels for eleven operational taxonomic units (OTUs) on days 1, 3, 7, 14, and 21 for each of the three mice. We eliminate any OTU that was not present in a measurable amount at all time points for any of the mice, which reduces our data to seven OTUs labeled \textit{Blautia}, \textit{Barnesiella}, \textit{Unclassified Mollicutes}, \textit{Undefined Lachnospiraceae}, \textit{Unclassified Lachnospiraceae}, \textit{Coprobacillus}, and \textit{Other}. Here, \textit{Other} is the eleventh original OTU and is the collection of bacteria not assigned to any of the other ten original OTUs. 

For our method we use all 21 (seven OTUs for three mice) time-series as the data, but we model the seven OTU interactions using a single, seven state Lotka-Volterra model given by
\begin{equation*}\label{eq:lvmicro}
	\bfy' = \bff(\bfy) = \diag{\bfy} (\bfb + \bfA \bfy), \ \bfy(1) = \bfy_0.
\end{equation*}
This then means that the model given to the objective function is actually
\begin{equation*}\label{eq:lvstack}
	\tilde\bfy' = \tilde\bff(\tilde\bfy) = \diag{\tilde\bfy} (\tilde\bfb + \tilde\bfA \tilde\bfy), \ \tilde\bfy(1) = \tilde\bfy_0
\end{equation*}
with
\begin{equation*}
\tilde\bfy = \begin{bmatrix*}[r] \bfy^1 \\[1ex] \bfy^2 \\[1ex] \bfy^3 \end{bmatrix*}, \ \tilde\bfb = \begin{bmatrix*}[r] \bfb \\[1ex] \bfb \\[1ex] \bfb \end{bmatrix*}, \ \tilde\bfA = \begin{bmatrix*}[r] \bfA & & \\[1ex] & \bfA & \\[1ex] & & \bfA \end{bmatrix*}, \ \tilde\bfy_0 = \begin{bmatrix*} \bfy^1_0 \\[1ex] \bfy^2_0 \\[1ex] \bfy^3_0 \end{bmatrix*}.
\end{equation*}
The superscripts indicate the data separately collected from each of the three mice, so our method approximates 21 different state variables, corresponding to the 21 total time-series, with spline functions. Those approximations, however, are all governed by a single seven-state Lotka-Volterra system defined by the intrinsic growth vector $\bfb$ and the interaction matrix $\bfA$.

The projection $\bfm$ in the objective function is the identity projection, and the weight matrix $\bfW$ is defined as $\bfW = \bfI_5 \otimes \diag{[w_1,\dots,w_{21}]\t}$ where $\bfI_5$ is the $5\times5$ identity matrix and $w_i = 1/(100\times\sigma(\bfD_i))$, $i = 1,\ldots,21$. We also replace the standard constraint term in the objective function with a sparsity constraint on the interaction matrix $\bfA$, which was defined earlier. To find the regularization parameters $\lambda$ and $\alpha$, we sample 100 $(\lambda, \alpha)$-pairs from the square $[1, 100] \times [10^{-6}, 10^{-4}]$ and choose a pair using 12-fold cross-validation. The $(\lambda, \alpha)$-pair is approximately $(2.6727, 5.7508 \times 10^{-6})$.

We then sample the parameter space 1,000 times using a Latin hypercube sampling and use these samples as initial parameter sets for local optimizations by the Gauss-Newton method.  The global minimizer is the local minimizer that most minimizes the objective function.

\section{Matlab Implementation}\label{apx:matlab} % (fold)
\label{sec:matlab_implementation}

\lstset{language=Matlab,%
    %basicstyle=\color{red},
    breaklines=true,%
    morekeywords={matlab2tikz},
    keywordstyle=\color{blue},%
    morekeywords=[2]{1}, keywordstyle=[2]{\color{black}},
    identifierstyle=\color{black},%
    stringstyle=\color{mylilas},
    commentstyle=\color{mygreen},%
    showstringspaces=false,%without this there will be a symbol in the places where there is a space
    numbers=left,%
    numberstyle={\tiny \color{black}},% size of the numbers
    numbersep=9pt, % this defines how far the numbers are from the text
    emph=[1]{for,end,break},emphstyle=[1]\color{blue}, %some words to emphasise
    %emph=[2]{word1,word2}, emphstyle=[2]{style},    
}

\subsection{Continuous Shooting} % (fold)
\label{sub:continuous_shooting}
\begin{lstlisting}
function [varargout] = continuousShooting(mFcn, nf, t, d, w, p, paramPE)
%
% function [optResults] = continuousShooting(mFcn, nf, t, d, w, p, paramPE)
%
% Author:
%   (c) Matthias Chung (mcchung@vt.edu)
%       Justin Krueger (kruegej2@vt.edu)
%
% Date: June 2015
%
% MATLAB Version: 8.4.0.150421 (R2014b)
%
% Description:
%   This parameter estimation procedure minimizes
%
%      min_(p,q) f(p,q) = a*||w(prjFcn(s(tau,q,t))-d)||^2 + b*lambda*||s'(tau,q,T)- mFcn(T,s(tau,q,T),p)||^2,
%
%   for a given ODE model function mFcn, parameter set p, a data set (t,d) and
%   a weight matrix w. lambda is a regularization on the accuracy of the
%   model. s is a cubic spline function uniquely determined by the
%   parameters q and knots tau. s(tau,q,t) means spline with parameters
%   q and knots tau evaluated at times t. s' is the time derivative
%   of spline s.
%
%   By default the method uses a Gauss-Newton optimization method and
%   requires the derivatives of f with respect to y and p.
%   Inital guess for y is d. Parameter a is set to 1/nd where nd is the 
%   total number of data points over all states. The parameter b is set to 
%   1/(nT*nf) where nT is the number of sample times T and nf is the 
%   dimension of the model.
%
%   This algorithm requires the function cubicSpline.m and possibly
%   others depending on the outputs requested.
%
% Input arguments:
%   mFcn        - model function of ODE y' = mFcn(t,y,p) of dimension nf
%   nf          - dimension of the model function
%   t           - time points of dimension 1 x n+1 where measurements are taken (or cell)
%   d           - data values at times t with dimension m x n+1 (or cell)
%   w           - weighting matrix for data values with dimension m x n+1 (or cell)
%   p           - inital guess for parameter values
%   #paramPE
%       optFcn   	- optimization function to be used [default @gaussNewtonJacobian]
%       prjFcn      - projection function of model onto observation (data requires derivatives of prjFcn) [default @linearProjection]
%       lambda      - regularization parameter (accuracy of model equations) [default 1]
%       ntau        - number of time points used to create initial guess for spline parameters q [default 50]
%       nT          - number of time points used to discretize model misfit norm [default 100]
%       regFcn      - additional regularization terms [default @regOrganizer]
%       alpha       - regularization parameters for model parameter regularization (row vector) [default 0]
%       beta        - regularization parameters for spline parameterregularization (row vector) [default 0]
%       q           - initial guess of model solution at time points tau [default does not exist]
%       paramOPT    - all typical parameter to be set for the TIA optimization toolbox [default {}]
%
% Output arguments:
%   varargout   
%       {1} - structure containing basic information such as minimizers, minimum objective function, etc.
%
% Example:
%   modelFcn = @lotkaVolterra;
%   nf = 4;
%   t = linspace(0, 2 * pi, 20);
%   d = [cos(t); sin(t)];
%   w = 10*bsxfun(@rdivide, ones(size(times)), std(data,0,2));
%   p = [2 2 3 1 -1 2]';
%   paramPE = {'lambda', 1e-1};
%   [optResults] = continuousShooting(modelFcn, nf, t, d, w, p, paramPE);
%
% References:
%

% convert data to appropriate format
if iscell(t)
    [t, d, w] = rearrangeData(t, d, w); % reshape if non-consistant measuring time points
end

% dimensions of parameter p
np = size(p, 1);

% set default parameters
optFcn = @gaussNewtonJacobian;                      % optimization algorithm
prjFcn = @linearProjection;                         % projection onto data
lambda = 1;                                         % regularization parameter for accuracy of model equations
ntau = 50;                                          % number of knots for spline
nT = 100;                                           % number of evalauation points for model misfit

% set default regularization for model and spline parameters (none)
regFcn = @regOrganizer; % regFcn will return 0's when used
alpha = 0;              % regularization parameter(s) for model regularization
beta = 0;               % regularization parameter(s) for spline regularization

% default options for optimization toolbox
paramOPT = {};

% rewrite default options if needed
if nargin == nargin(mfilename)
    for j = 1:size(paramPE,1)
        eval([paramPE{j,1},'= paramPE{j,2};']);
    end
end

% set outputs of misfit function to be either residual or fuction value
res = 1;
if ~strcmp(func2str(optFcn), 'gaussNewtonJacobian')
    res = 0;
end

% time discretizations
tau = linspace(t(1), t(end), ntau);     % set default knots of the spline
T = linspace(t(1), t(end), nT);         % choose default evaluation points of T for model misfit s'(tau,q,T) - f(T,s(tau,q,T),p)

% inital guess for spline parameters if none is given
if ~exist('q','var')
    q = zeros(nf, ntau);
    % construct q for one state variable at a time
    for j = 1:nf
        idxData = find(d(j,:) ~= 0);
        % ensure at least two data points for spline
        if length(idxData) > 2 
            q(j,:) = cubicSpline(t(idxData), d(j,idxData), tau);    % interpolate data for q values
        end
    end
end

% reshape w and d
w = w(:);
d = d(:);

% remove all indices where no data is available
idxFull = find(w ~= -1);
nidxFull = length(idxFull);

% remove all indices where no data is available and all indices the user wishes to ignore
idx = find(w ~= -1 & w ~= 0);
nidx = length(idx);
w = w(idx)/sqrt(nidxFull); % normalization of residual weights w in s(tau,q,t)-d by sqrt(nd)
d = d(idx);

% pre-compute dsdqD, dsdqM, and dsdqdtM
[~, ~, dsdqD]          = cubicSpline(tau, q, t);    % get spline derivatives for data misfit
[~, ~, dsdqM, dsdqdtM] = cubicSpline(tau, q, T);    % get spline derivatives for model misfit
e = speye(nf, nf);
dSdQD = kron(dsdqD, e);                             % replicate dsdqD for correct model dimension
dSdQM = kron(dsdqM, e);                             % replicate dsdqM for correct model dimension
dSdQdtM = kron(dsdqdtM, e);                         % replicate dsdqdtM for correct model dimension

% objective function
oFcn = @(pq) misfitFcn(pq, mFcn, nf, d, w, np, prjFcn, idx, nidx, lambda, tau, ntau, T, nT, dsdqD, dsdqM, dsdqdtM, dSdQD, dSdQM, dSdQdtM, res, regFcn, alpha, beta);

% initial search parameters pq = (p (unknown),q)
pq = [p; q(:)];

% solve parameter estimation scheme
[pq, f] = optFcn(oFcn, pq, paramOPT);

% extract optimal parameters p and q, approximated state y, and initial condition y0
p = pq(1:np);                               % update optimized parameters
q = reshape(pq(np+1:end), nf, ntau);                   % reshape q so the spline parameter set has the proper dimensions
y = cubicSpline(tau, q, t);                             % construct an approximation of states y using a cubic spline and optimized spline parameters
y0 = y(:, 1);                                           % provide the optimized initial conditions for the states y

% store optimization and decomposition results
varargout{1} = cell2struct({p, q, y0, y, f}, {'pMin', 'qMin', 'y0Min', 'yMin', 'fMin'}, 2);

end

% -------------------------------------------------------------------------
% arrange times, data, and weights in a usuable form
function [t, d, w] = rearrangeData(T, D, W)

% create sorted vector of measuring times and remove duplicate entries
t = unique(sort(cell2mat(T')));

% generate -1 at non measured points in d and w
w = -1*ones(length(T), length(t));
d = -1*ones(length(T), length(t));
for j = 1:length(T)
    for i = 1:length(T{j})
        w(j,T{j}(i) == t) = W{j}(i); % replace -1 any place a weight is given
        d(j,T{j}(i) == t) = D{j}(i); % replace -1 any place a data point is given
    end
end

end

% -------------------------------------------------------------------------
% calculate the objective function value and/or other needed values
function [varargout] = misfitFcn(pq, mFcn, nf, d, w, np, prjFcn, idx, nidx, lambda, tau, ntau, T, nT, dsdqD, dsdqM, dsdqdtM, dSdQD, dSdQM, dSdQdtM, res, regFcn, alpha, beta)

% split search parameter pq into parameters p and q
p = pq(1:np);                               % update optimized parameters
q = reshape(pq(np+1:end), nf, ntau);        % reshape q so the spline parameter set has the proper dimensions

% normalization of residuals s'(T)- f(T,s(tau,q,T),p)
lambda = sqrt(lambda/(nf*nT));

% calculate spline interpolations and derivative of splines
sD = q*dsdqD';
sM = q*dsdqM';
dsdtM = q*dsdqdtM';

% [sD]        = cubicSpline(tau, q, t); % evaluate s(tau,q,t) for data misfit
% [sM, dsdtM] = cubicSpline(tau, q, T); % evaluate s(tau,q,T) for model misfit

% only r or f is required
if nargout == 1
    
    % calculate model function f
    F = mFcn(T, sM, p);
    
    % calculate projection
    prjsD = prjFcn(sD); % calculate projection
    prjsD = prjsD(:);   % vectorize projection
    prjsD = prjsD(idx); % remove indices of unused values from the projection
    
    % residual
    r = [w.*(prjsD-d); lambda*(dsdtM(:)-F(:))]; % combined residual of data and model misfit
    
    % r is required
    if res 
        rR = regFcn(p, tau, q, res, alpha, beta);    % calculate residual for regularization term(s)
        varargout{1} = [r; rR];                             % add regularization residuals
    % f is required
    else
        f = regFcn(p, tau, q, res, alpha, beta); % calculate function value for regularization term(s)
        varargout{1} = 0.5*(r'*r) + f;                  % calculate function
    end

% more than r or f is required
else
    
    % calculate model function f and derivatives fs and fp
    [F, Fs, Fp] = mFcn(T, sM, p);
   
    % calculate projection
    [prjsD, dprjsD] = prjFcn(sD);   % calculate projection and its derivative
    prjsD = prjsD(:);               % vectorize projection
    prjsD = prjsD(idx);             % remove indices of unused values from the projection
    dprjsD = dprjsD(idx,:);         % remove rows of derivatives due to unused values
    
    % residual and Jacobian
    r = [w.*(prjsD-d); lambda*(dsdtM(:)-F(:))];                                                             % combined residual of data and model misfit
    J = [sparse(nidx,np), spdiags(w,0,nidx,nidx)*dprjsD*dSdQD; -lambda*Fp, lambda*(dSdQdtM - Fs*dSdQM)];   % Jacobian of data and model misfit
    
    % r and J are required
    if res 
        [rR, JR] = regFcn(p, tau, q, res, alpha, beta);  % calculate residual and Jacobian for regularization term(s)
        varargout{1} = [r; rR];                                 % add regularization residuals
        varargout{2} = [J; JR];                                 % add regularization Jacobian
    % more than f is required
    else
        % f and g are required
        if nargout == 2
            [f, g] = regFcn(p, tau, q, res, alpha, beta);    % calculate function and gradient values for regularization term(s)
            varargout{1} = 0.5*(r'*r) + f;                          % calculate function
            varargout{2} = J'*r + g;                                % calculate gradient 
        % f, g, and H are required
        else 
            [f, g, H] = regFcn(p, tau, q, res, alpha, beta); % calculate function, gradient, and Hessian values for regularization term(s)
            varargout{1} = 0.5*(r'*r) + f;                          % calculate function 
            varargout{2} = J'*r + g;                                % calculate gradient
            varargout{3} = J'*J + H;                                % Calculate Hessian
        end
    end
    
end

end
\end{lstlisting}
% subsection continuous_shooting (end)

\subsection{Cubic Splines} % (fold)
\label{sub:spline_function}
\begin{lstlisting}
function [s, dsdt, dsdy, dsdydt] = cubicSpline(t, y, tt)
%
% function [s, dsdt, dsdy, dsdydt] = cubicSpline(t, y, tt)
%
% Author:
%   (c) Matthias Chung (mcchung@vt.edu)
%       Justin Krueger (kruegej2@vt.edu)
%
% Date: February 2014
%
% MATLAB Version: 8.1.0.604 (R2013a)
%
% Description:
%   This function calculates the piecewise polynomials (cubic spline) or the values of
%   a cubic spline s with given knots t and corresponding data y(t). If the
%   third argument tt is given this function evaluates the piecewise polynomials
%   at timepoints tt. Not-a-knot boundary condition are used.
%   Additionally this function calculates derivatives ds/dt, ds/dy and d/dy(ds/dt).
%   Note, the derivatives ds/dy and d/dy(ds/dt) are independent of y and for
%   multiple dimensions of y, dsdy and d/dy(ds/dt) are of the form
%     dsdy = [B 0 ... 0 0]
%            [: :     : :]
%            [0 0 ... 0 B]
%   This function just returns just the diagonal block B for efficiency.
%
% Input arguments:
%   t   - knots of spline (row vector 1 x n+1)
%   y   - interpolation values of spline (matrix m x n+1)
%   tt  - evaluation points of spline (row vector 1 x k)
%
% Output arguments:
%   s        - spline s (piecewise polynomial, evaluated at tt if tt not empty)
%   dsdt     - derivative dsdt of s with respect to t (piecewise polynomial, evaluated at tt if tt not empty)
%   dsdy     - derivative dsdy of s with respect to y (piecewise polynomial, evaluated at tt if tt not empty)
%   dsdydt   - derivative dsdydt of s with respect to y and t (piecewise polynomial, evaluated at tt if tt not empty)
%
% Example:
%   t = linspace(0, 2 * pi, 20);
%   y = [cos(t); sin(t)];
%   tt = linspace(0, 2 * pi, 100);
%   [s, dsdt, dsdy, dsdydt] = cubicSpline(t, y); (returns piecewise polynomials)
%   [s, dsdt, dsdy, dsdydt] = cubicSpline(t, y, tt); (returns spline values)
%
% References:
%   [1] Carl DeBoor, A Practical Guide to Splines, Reprint edition, Springer, 1994.
%

% initialize fixed options of algorithm
n = length(t)-1;

% check input data
if n < 1
    error('There must be at least two data points.');
end

% initialize constants
h = diff(t);         % length of interval parts h
m = size(y, 1);       % dimension of y
diffy = diff(y, 1, 2); % difference of interpolation points

if n == 1 % interpolant is a straight line
    
    s = mkpp(t, [diffy/h, y(:,1)], m);
    if nargout > 1
        dsdt = mkpp(t, diffy/h, m);
        if nargout > 2
            dsdy = mkpp(t, [-1, h; 1, 0 ]/h, 2)';
            if nargout == 4
                dsdydt = mkpp(t, [-1, 1]/h, 2)';
            end
        end
    end
    
elseif n == 2 % interpolant is a parabola
    
    a = (diffy(:,2)/h(2) - diffy(:,1)/h(1))/(h(1)+h(2)); % quadratic coefficient
    s = mkpp(t([1,3]), [a, diffy(:,1)/h(1) - a*h(1), y(:,1)], m); % construct spline s
    if nargout > 1 % construct dsdt
        dsdt = mkpp(t([1,3]), [2*a, diffy(:,1)/h(1) - a*h(1)], m);
        if nargout > 2 % construct dsdy
            a = 1/(h(1)+h(2)) * [1/h(1); -1/h(2)-1/h(1); 1/h(2)]; % quadradic coefficient
            dsdy = mkpp(t([1,3]),[a, 1/h(1)*[-1; 1; 0] - h(1)*a, [1; 0; 0] ], 3)';
            if nargout == 4 % construct dsdydt
                dsdydt = mkpp(t([1,3]), [2*a, 1/h(1)*[-1; 1; 0] - h(1)*a], 3)';
            end
        end
    end
    
else % interpolant is a regular spline
    
    % upper diagonal of matrix
    lambda = (h(2:n)./(h(1:n-1)+h(2:n)))';
    
    % initialize tri-diagonal matrix
    A = spdiags([[1-lambda; 0; 0], 2*ones(n+1,1), [0; 0; lambda]], -1:1, n+1, n+1);
    
    % initialize right hand side of linear system
    B = [zeros(m,1), 6*(bsxfun(@rdivide,diffy(:,2:n),(h(1:n-1)+h(2:n)).*h(2:n)) - bsxfun(@rdivide,diffy(:,1:n-1),(h(1:n-1)+h(2:n)).*h(1:n-1))), zeros(m,1)]';
    
    % set not-a-knot boundary conditions
    A(1, 1:3) = [ -1/h(1), 1/h(1)+1/h(2), -1/h(2)];
    A(n+1, n-1:n+1) = [-1/h(n-1), 1/h(n-1)+1/h(n), -1/h(n)];
    
    M =  (A\B)'; % solve for moments M
    H = bsxfun(@times, ones(m,1), h); % distances h for each dimension of y
    
    % generate piecewise polynomial
    s = mkpp(t, [diff(M,1,2)./(6*H), M(:,1:n)/2, diffy./H - (2*M(:,1:n)+M(:,2:n+1)).*H/6, y(:,1:n)], m);
    
    if nargout > 1 % calculate derivative dsdt
        
        dsdt = mkpp(t, [diff(M,1,2)./(2*H), M(:,1:n), diffy./H - (2*M(:,1:n)+M(:,2:n+1)).*H/6], m); % calculate dsdt
        
        if nargout > 2 % calculate derivatives dsdy and dsdydt
            
            % initialize Jacobian of right hand side
            mu = -6./(h(1:n-1).*h(2:n))';
            JB = spdiags([[-mu.*lambda; 0; 0], [0; mu; 0], [0; 0; -mu.*(1-lambda)]], -1:1, n+1, n+1);
            
            % calculate Jacobian of the moments
            JM = full(A\JB);
            
            % initialize constants
            H = bsxfun(@times, h', ones(1,n+1));
            I0 = [sparse(n,1), speye(n)];
            IN = speye(n, n+1);
            
            % calculate coefficients
            Jc = 1./H.*(I0-IN) - H/6.*(2*IN+I0)*JM;
            Jb = 1/2*IN*JM;
            Ja = 1./(6*H).*(I0-IN)*JM;
            
            dsdy = mkpp(t, [Ja; Jb; Jc; full(IN)]', n+1); % calculate dsdy
            
            if nargout == 4 % calculate dsdydt
                dsdydt = mkpp(t, [3*Ja; 2*Jb; Jc]', n+1);
            end
        end
    end
    
end

% evaluate piecewise polynomial at time points tt otherwise return piecwise polynomials
if nargin > 2 % return spline values
    s = ppval(s, tt);
    if nargout > 1
        dsdt = ppval(dsdt, tt);
        if nargout > 2
            dsdy = ppval(dsdy, tt)';
            if nargout == 4
                dsdydt = ppval(dsdydt, tt)';
            end
        end
    end
end

end
\end{lstlisting}
% subsection spline_function (end)

\subsection{Lotka-Volterra Model} % (fold)
\label{sub:lotka_volterra_model}
\begin{lstlisting}
function [f, dfdy, dfdp] = lotkaVolterra(~, y, p)
%
% function [f, dfdy, dfdp] = lotkaVolterra(~, y, p)
%
% Author:
%   (c) Matthias Chung (mcchung@vt.edu)
%       Justin Krueger (kruegej2@vt.edu)
%
% Date: July 2014
%
% MATLAB Version: 8.1.0.604 (R2013a)
%
% Description:
%   The Lotka-Volterra differential equation
%     y' =  diag(y)(r+Ay)
%   with nf species (length of r is nf and dimension of A is nf x nf).
%   A and r are collected in the parameter p = [r; vec(A)].
%
% Input arguments:
%   ~        - time (no time dependence, autonomous system)
%   y        - value at time(s) t
%   p        - parameters (length nf^2+nf)
%
% Output arguments:
%   f        - first output argument is the model function
%                  f = [y_1';...;y_nf']
%   dfdy     - second output argument is df/dy
%   dfdp     - third output argument is df/dp
%
%   f    = [f_1(t_1)  ... f_1(t_nt) ]
%          [  ...     ...    ...    ]
%          [f_nf(t_1) ... f_nf(t_nt)]
%
% Dimension: (nf*nt) x (nf*nt)
%   dfdy = [df/dy(t_1)   0   ...   0]
%          [0 ...     ...     ...  0]
%          [0   ...  0   df/dy(t_nt)]
%     
%   where df/dy(t) = [df_1/dy_1(t)   df_1/dy_2(t)  ... df_1/dy_nf(t) ]
%                    [   ...             ...       ...      ...      ]
%                    [df_nf/dy_1(t)  df_nf/dy_2(t) ... df_nf/dy_nf(t)]
%
% Dimension: (nf*nt) x (np)
%   dfdp = [df/dp(t_1) ]
%          [    ...    ]
%          [df/dp(t_nt)]
%
%   where df/dp(t) = [df_1/dp_1(t)  df_1/dp_2(t)  ... df_1/dp_np(t) ]
%                    [   ...            ...       ...       ...     ]
%                    [df_nf/dp_1(t) df_nf/dp_2(t) ... df_nf/dp_np(t)]
% 
%   and p = [r; vec(A)]
%
% To expand: use objects to calculate dfdp and dfdy.
%
% Example:
%   [f, dfdy, dfdp] = lotkaVolterra(~, [4 6]', [2 2 3 1 -1 2]')
%
% References:
%

% record dimensions of y
[nf, nt] = size(y);

% decompose parameters
r = p(1:nf);                        % intrinsic growth vector
A = reshape(p(nf+1:end), nf, nf);   % interaction matrix

% model function
f = DYDT(y, r, A, nf, nt);

% df/dy required
if nargout > 1
    dfdy = DFDY(y, r, A, nf, nt); % derivative with respect to the state
    % df/dp required
    if nargout > 2
        dfdp = DFDP(y, nf, nt); % derivative with respect to the parameters
    end
end

end

% -------------------------------------------------------------------------
% Lokta-Volterra model f
function f = DYDT(y, r, A, nf, nt)
f = y.*(bsxfun(@times,r,ones(nf,nt)) + A*y);
end

% -------------------------------------------------------------------------
% derivative of f respect to y
function dfdy = DFDY(y, r, A, nf, nt)
m = nf*nt;
dfdy = spdiags(repmat(r,nt,1),0,m,m) + spdiags(y(:),0,m,m)*kron(speye(nt,nt),A) + spdiags(reshape(A*y,m,1),0,m,m);
end

% -------------------------------------------------------------------------
% derivative of f respect to p
function dfdp = DFDP(y, nf, nt)
m = nf*nt;
dfdr = spdiags(y(:),0,m,m)*repmat(speye(nf,nf),nt,1);
dfdA = spdiags(y(:),0,m,m)*kron(y',speye(nf,nf));
dfdp = [dfdr, dfdA];
end
\end{lstlisting}
% subsection lotka_volterra_model (end)

% \subsection{Simulation Study Example} % (fold)
% \lstinputlisting{../code/exampleScript.m}
% % subsection spline_function (end)
 
% \subsection{Gauss Newton Optimization} % (fold)
% \label{sub:gauss_newton_optimization}
% \lstinputlisting{../code/gaussNewton.m}
% % subsection gauss_newton_optimization (end)
%
% \subsection{Strong Wolfe Line Search} % (fold)
% \label{sub:strong_wolfe_line_search}
% \lstinputlisting{../code/strongWolfe.m}
% % subsection strong_wolfe_line_search (end)
%
% \subsection{Data Projection Function} % (fold)
% \label{sub:projection_function}
% \lstinputlisting{../code/linearProjection.m}
% % subsection projection_function (end)
%
% \subsection{Regularization Organizer} % (fold)
% \label{sub:regularization_organizer}
% \lstinputlisting{../code/regOrganizer.m}
% % subsection regularization_organizer (end)
%
% \subsection{Sparsity Regularization} % (fold)
% \label{sub:sparsity_regularization}
% \lstinputlisting{../code/sparsityRegLV.m}
% % subsection sparsity_regularization (end)
%
% \subsection{Huber Function} % (fold)
% \label{sub:huber_function}
% \lstinputlisting{../code/huberFcn.m}
% % subsection huber_function (end)

% section matlab_implementation (end)

\end{document}